\documentclass[cernpreprint, texlive=2020, UKenglish, texmf, orcidlogo]{atlasdoc}
 
\usepackage{atlaspackage}
\usepackage{atlasbiblatex}
\usepackage{atlasphysics}
\usepackage{verbatim}

\newcommand{\PbPb}{\mbox{Pb+Pb}}
\newcommand{\pbpb}{\mbox{Pb+Pb}}
\newcommand{\dR}{\mbox{$\Delta R$}}
 
\newcommand{\bhadrons}{\mbox{$b$-hadrons}}

\newcommand{\bhadron}{\mbox{$b$-hadron}}

\newcommand{\bmesons}{\mbox{$B$-mesons}}
\newcommand{\dmeson}{\mbox{$D$-meson}}

\newcommand{\Ncoll}{\mbox{$N_{\mathrm{coll}}$}}

\newcommand{\Njet}{\mbox{$N_{\mathrm{jet}}$}}

\newcommand{\Npart}{\mbox{$N_{\mathrm{part}}$}}

\newcommand{\pythiaeight}{\textsc{Pythia}\,8}             
\newcommand{\herwig}{{\textsc{Herwig++}}}

\newcommand{\RAA}{\mbox{$R_{\mathrm{AA}}$}}
\newcommand{\raa}{\mbox{$R_{\mathrm{AA}}$}}
\newcommand{\RFive}{\mbox{$R = 0.5$}}
\newcommand{\RFour}{\mbox{$R = 0.4$}}

\newcommand{\RTwo}{\mbox{$R= 0.2$}}

\newcommand{\TAA}{\mbox{$T_{\mathrm{AA}}$}}

\newcommand{\invnb}{\mbox{${\mathrm{nb}^{-1}}$}}
\newcommand{\invpb}{\mbox{${\mathrm{pb}^{-1}}$}}

\newcommand{\pTjet}{\mbox{$p_{{\mathrm{T}}}$}}

\newcommand{\pTtrk}{\mbox{$p_{{\mathrm{T}}}^{\mathrm{ch}}$}}

\newcommand{\pTtrue}{\mbox{$p_{\mathrm{T}}^{\mathrm{truth}}$}}

\newcommand{\ETjet}{\mbox{$\ET^{\mathrm{jet}}$}}

\newcommand{\ptjet}{\mbox{$p_{\mathrm{T}}^{\mathrm{jet}}$}}

\newcommand{\ptrecojet}{\mbox{$p_{\mathrm{T, reco}}^\textrm{jet}$}}
\newcommand{\ptrecojetmu}{\mbox{$p_{\mathrm{T, reco}}^{\textrm{jet}+\mu}$}}

\newcommand{\cjets}{\textit{c}-jets}
\newcommand{\cjet}{\textit{c}-jet}
\newcommand{\ljets}{\textit{light}-jets}
\newcommand{\ljet}{\textit{light}-jet}

\newcommand{\chadron}{\textit{c}-hadron}
\newcommand{\jetmu}{{jet}+$\mu$}
\newcommand{\ptrel}{\mbox{$p_\mathrm{T}^\mathrm{rel}$}}
\newcommand{\z}{\mbox{$z$}}
\newcommand{\Fbkg}{\mbox{$F_{\mathrm{mix},i}$}}
\newcommand{\Fb}{\mbox{$F_{\mathrm{b},i}$}}
\newcommand{\Fc}{\mbox{$F_{\mathrm{c},i}$}}
\newcommand{\ith}{\mbox{\textit{i}-th}}
\newcommand{\Fl}{\mbox{$F_{\mathrm{l},i}$}}
\newcommand{\fbkg}{\mbox{$f_{\mathrm{mix},i}$}}

\newcommand{\fc}{\mbox{$f_{\mathrm{c},i}$}}
\newcommand{\fb}{\mbox{$f_{\mathrm{b},i}$}}

\newcommand{\ptmu}{\mbox{$p_{{\mathrm{T}}}^{\mu}$}}

\graphicspath{{logos/}{figures/}}

\AtlasTitle{Measurement of the nuclear modification factor of $b$-jets in 5.02~TeV Pb+Pb
collisions with the ATLAS detector}
 
\PreprintIdNumber{CERN-EP-2022-043}
\AtlasAbstract{
This paper presents a measurement of $b$-jet production in Pb+Pb and $pp$
collisions at $\sqrt{s_{_\textrm{NN}}}=5.02$~TeV with the ATLAS detector at the LHC.
The measurement uses 260~$\textrm{pb}^{-1}$ of $pp$ collisions collected in 2017
and 1.4~$\textrm{nb}^{-1}$ of Pb+Pb collisions collected in 2018.
In both collision systems, jets are reconstructed via the anti-$k_{t}$ algorithm.
The $b$-jets are identified
from a sample of jets containing muons from the semileptonic decay of
$b$-quarks using template fits of the muon momentum relative to the jet
axis.
In $pp$ collisions, $b$-jets are reconstructed
for radius parameters $R= 0.2$ and $R= 0.4$, and only $R= 0.2$ jets are used  in Pb+Pb collisions.
For comparison, inclusive $R= 0.2$ jets are also measured
using 1.7~$\textrm{nb}^{-1}$ of Pb+Pb collisions collected
in 2018 and the same $pp$ collision data as the $b$-jet measurement.
The nuclear modification factor, $R_\textrm{AA}$, is calculated for both
$b$-jets and inclusive jets with $R= 0.2$ over the
transverse momentum range of 80--290~GeV.
The nuclear modification factor for $b$-jets decreases from peripheral
to central collisions.
The ratio of the $b$-jet $R_\textrm{AA}$
to inclusive jet $R_\textrm{AA}$ is also presented and
suggests that the $R_\textrm{AA}$ for $b$-jets is larger than that for inclusive jets
in central Pb+Pb collisions.
The measurements are compared with theoretical calculations and suggest a role for mass and colour-charge effects in partonic energy loss in heavy-ion collisions.
}
 
\author{The ATLAS Collaboration}
 
\AtlasJournalRef{Eur. Phys. J. C (2023) 83, 438 }
\AtlasDOI{10.1140/epjc/s10052-023-11427-9}

\hypersetup{pdftitle={ATLAS document},pdfauthor={The ATLAS Collaboration}}
 
\begin{document}
 
\maketitle
 
\tableofcontents

\section{Introduction}
\label{sec:intro}

% The next lines are included from the .//chapters/Introduction.tex input file
\newcommand{\AtlasCoordFootnote}{
ATLAS uses a right-handed coordinate system with its origin at the nominal interaction point (IP)
in the center of the detector and the \(z\)-axis along the beam pipe.
The \(x\)-axis points from the IP to the center of the LHC ring,
and the \(y\)-axis points upwards.
Cylindrical coordinates \((r,\phi)\) are used in the transverse plane,
\(\phi\) being the azimuthal angle around the \(z\)-axis.
The pseudorapidity is defined in terms of the polar angle \(\theta\) as \(\eta = -\ln \tan(\theta/2)\).
Angular distance is measured in units of \(\Delta R \equiv \sqrt{(\Delta\eta)^{2} + (\Delta\phi)^{2}}\).}
 
The primary physics aim of the heavy-ion programme at the Large Hadron Collider (LHC) is to produce
and study the quark--gluon plasma (QGP), the high-temperature state of quantum-chromodynamic (QCD) matter
in which quarks and gluons are no longer confined within protons and neutrons (for a recent review, see
Ref.~\cite{Busza:2018rrf}).
Measurements of jets arising from hard-scattered partons produced in the early stages of
heavy-ion collisions provide information about the short-distance-scale
interactions of high-energy partons with the QGP.  The overall rate of jets in central Pb+Pb collisions at a given transverse momentum,
\pt, is found to be significantly lower than expectations based on
\pp\ collisions, up to a \pt of approximately 1~\TeV~\cite{Aaboud:2018twu,Acharya:2019jyg,CMS:2021vui}.
The showering partons which give rise to jets
undergo medium-induced gluon radiation and
elastic scattering off the QGP constituents; both processes
broaden the angular distribution of the energy from the parton shower~\cite{ATLAS:2019pid}.
This leads
to the measurement of lower-energy jets  than in \pp\ collisions as some of the parton's energy is moved
outside of the jet cone.
 
The interactions between high-energy partons
and the QGP are expected to depend on the parton's QCD colour charge and
mass. Thus,
jets originating from \bquarks, \bjets,
(for a recent review, see Ref.~\cite{Dong:2019byy})
are of particular interest because the quark mass is large
compared to light quarks. Additionally, the QCD colour charge is controlled;
this contrasts with inclusive jets, which
mostly originate from  a mixture of light quarks and gluons.
The \bjets\ produced in hadronic collisions  consist  of jets in which the \bquark\
is directly produced in the hard scattering (e.g.\ $gg \to b\bar{b}$)
and jets in which a hard-scattered gluon splits into a $b\bar{b}$ pair ($g \to b\bar{b}$) as part of the
showering process.
 
Inclusive jets and \bjets\ are expected to be sensitive to
different effects in the QGP.
Medium-induced gluon radiation is expected to be
suppressed for heavy quarks by the dead-cone effect~\cite{Dokshitzer:2001zm}.
As the velocity of the \bquark\ increases
the importance of the dead-cone effect decreases.
Calculations for \pbpb\ collisions
show that the dead-cone effect reduces the suppression of \bmesons\
up to transverse momenta of approximately 50~\GeV~\cite{Zigic:2018ovr}.
This does not simply translate to a specific jet \pt\ range because
\bquarks\ in jets from gluon splitting tend to have lower momentum
than those directly from hard scattering at the same jet \pt.
Differences in the interactions between the QGP and the
developing parton shower between \bjets\ and inclusive jets
could  also arise from
differences in how quarks and gluons interact with the QGP~\cite{Spousta:2015fca}.
Measurements of heavy-quark jets in
heavy-ion collisions are also expected to be sensitive to
the mixture of radiative and collisional energy loss in the
QGP~\cite{Wicks:2005gt, Ke:2018tsh, Dai:2018mhw}.
Due to these  different effects, which are expected to lead to
differences between \bjets\ and inclusive jets in \pbpb\ collisions,
it is important to compare the \bjet\ and inclusive-jet suppression
in \pbpb\ collisions over as wide a kinematic range as possible.
 
Prior measurements have investigated the effects of energy loss for heavy quarks.
Many measurements are based on electrons or muons from the
semileptonic decay of charm and bottom
hadrons~\cite{Adare:2006nq,Adare:2015hla,ALICE:2020sjb,ATLAS:2021xtw,STAR:2021uzu}.
Other measurements have partially~\cite{ATLAS:2018hqe,CMS:2012bms,CMS:2016mah,CMS:2018bwt}
or fully~\cite{CMS:2017qjw,CMS:2017uoy,STAR:2018zdy,ALICE:2021rxa,ALICE:2021kfc,ALICE:2021bib} reconstructed the heavy-flavour hadrons.
One limitation of these measurements is  that they do not
provide direct information about the total jet energy.
This is important because the fragmentation functions of \bjets\ are
very different from those of inclusive jets~\cite{ParticleDataGroup:2020ssz};  thus, it is
not possible to directly compare
the suppression of \bquarks\ with that of light quarks and gluons
without full jet measurements.
CMS has measured
the suppression of \bjets\ in \pbpb\ collisions at 2.76~\TeV\ and found it to be consistent
with that of inclusive jets over the jet transverse
momentum, \ptjet, range of 70--250~\GeV~\cite{Chatrchyan:2013exa},
although the uncertainties are large.
 
For the measurements reported in this article, jets are clustered with the \antikt\ algorithm~\cite{Cacciari:2008gp} using radius parameters
\RTwo\ (in both \pbpb\ and \pp\ collisions) and \RFour\ (only in \pp\ collisions).
At generator level, a jet is considered as a \bjet if a \bhadron\
with $\pt > 5$~\GeV\ is found within an angular distance\footnote{ATLAS uses a right-handed
coordinate system with its origin at the nominal interaction point (IP) in the centre of the detector, and the $z$-axis along the beam pipe.
The $x$-axis points from the IP to the centre of the LHC ring, and the $y$-axis points upward.
Cylindrical coordinates $(r,\phi)$ are used in the transverse plane, $\phi$ being the azimuthal angle around the $z$-axis.
The pseudorapidity is defined in terms of the polar angle $\theta$ as $\eta=-\ln\tan(\theta/2)$.
The rapidity is defined as $y = 0.5\ln[(E + p_z)/(E-p_z)]$ where $E$ and $p_z$ are the energy and $z$-component of the momentum along the beam direction, respectively.
Transverse momentum and transverse energy are defined as $\pt = p \sin\theta$ and $\ET = E \sin\theta$, respectively.
The angular distance between two objects with relative differences $\Delta \eta$ in pseudorapidity and $\Delta \phi$ in azimuth is given by $\Delta R = \sqrt{(\Delta \eta )^2 + (\Delta \phi)^2}$.}
of $\Delta R = 0.3$ from the jet axis.
The \bjets\ are identified statistically from a sample
of jets containing muons.
The muons from
\bhadron\ decays
are distinguished from muons arising from charm- and light-hadron decays by the difference between
their transverse momenta, \ptmu,
relative to the \jetmu\ axis. The \jetmu\ axis is defined in the transverse plane by
\begin{equation}
\vec{u}_\textrm{T}^{\textrm{jet}+\mu} = \frac{\vec p_\textrm{T}^{\mu}+\vec p_\textrm{T}^{\mathrm{jet}}}{\left|\vec p_\textrm{T}^{\mu}+\vec p_\textrm{T}^{\mathrm{jet}}\right|}
\label{eq:jetmuaxis}
\end{equation}
(where the vector notation indicates the vector quantity is meant instead of the magnitude of the vector), and the transverse momentum of the muon relative to the jet axis,
\ptrel, is defined as
\begin{equation}
\ptrel\ = \left| \vec p_\mathrm{T}^{\mu} \times \vec{u}_\textrm{T}^{\textrm{jet}+\mu} \right| .
\label{eq:ptrel}
\end{equation}
This method was used previously by ATLAS for \pp\ collisions at 7~\TeV\ and
the results were compared with those based on the reconstruction of secondary vertices~\cite{ATLAS:2011ac}.
Only  \RTwo\ jets are used in \pbpb\ collisions because they have better
angular resolution and have a smaller contribution from accidental jet--muon overlaps
in \pbpb\ collisions; both of these features improve the performance of the \ptrel\ method.
 
This paper reports measurements of inclusive jet and \bjet cross-sections in \pp collisions and
the per-event yield for inclusive jets and \bjets\ in \pbpb\ collisions at 5.02~\TeV.
In both collision  systems, the spectra are corrected to the generator level
(just before the \bhadron\ decays for \bjets).
For \RTwo\ jets the nuclear modification factor, \RAA, is calculated for \bjets\ as
\begin{equation*}
\raa^{\bjet} \equiv \left. \left.\frac{1}{N_\textrm{evt}} \frac{\textrm{d}^2 N_\textrm{AA}^{\bjet}}{\textrm{d}\pt\textrm{d}y}\right|_\textrm{cent} \middle/ \langle T_{\textrm{AA}}\rangle
\frac{\textrm{d}^{2}\sigma_{\pp}^{\bjet}}{\textrm{d}\pt \textrm{d}y} \right.
\label{eq:RAAb}
\end{equation*}
where $\textrm{d}^2 N_\textrm{AA}^{\bjet}/\textrm{d}\pt\textrm{d}y$ is the yield of \bjets\
in \pbpb\ collisions for the centrality,
\pt, and rapidity range of interest;
$\textrm{d}^{2}\sigma_{\pp}^{\bjet}/\textrm{d}\pt \textrm{d}y$
is the \bjet\ cross-section in \pp\ collisions;  $\langle T_{\textrm{AA}}\rangle$
is the nuclear thickness function~\cite{Miller:2007ri}; and
$N_\textrm{evt}$ is the number of  minimum-bias (MB) \pbpb\
events for the centrality selection under consideration.
Analogously, the nuclear modification factor for inclusive jets is defined as
\begin{equation*}
\raa^{\mathrm{inclusive \: jet}} \equiv \left. \left. \frac{1}{N_\textrm{evt}}
\frac{\textrm{d}^2 N_\textrm{AA}^{\mathrm{inclusive \: jet}}}{\textrm{d}\pt\textrm{d}y}\right|_\textrm{cent}
\middle/ \langle T_{\textrm{AA}}\rangle
\frac{\textrm{d}^{2}\sigma_{\pp}^{\mathrm{inclusive \: jet}}}{\textrm{d}\pt \textrm{d}y} \right.
\label{eq:RAAinc}
\end{equation*}
where the yield and cross-section are for inclusive jets and the other quantities
remain the same.

The measurements are performed in Pb+Pb collisions at $\sqrt{s_{_\textrm{NN}}}=5.02$~\TeV\ collected during 2018,
with an integrated luminosity of 1.4~$\textrm{nb}^{-1}$ for the \bjet sample and 1.7~$\textrm{nb}^{-1}$
for the inclusive jet sample,
and in \pp collisions at $\sqrt{s}=5.02$~\TeV\ collected during 2017,
with an integrated luminosity of 260~\invpb.
The measurements are presented for jets with $80<\pt<290$~\GeV\ and $|y|<2.1$.
 
The paper is structured as follows. Section~\ref{sec:detector} describes the ATLAS detector, and
Section~\ref{sec:selection} discusses the selection procedure applied to the data.
The data analysis is presented in Section~\ref{sec:analysis}
and the systematic uncertainties are presented in Section~\ref{sec:systematics}.
The results and a summary are presented in Sections~\ref{sec:result} and~\ref{sec:summary}.

% End of text imported from the .//chapters/Introduction.tex input file

\section{ATLAS detector}
\label{sec:detector}

% The next lines are included from the .//chapters/detector.tex input file
 
The ATLAS detector~\cite{PERF-2007-01} at the LHC covers nearly the entire solid angle around the collision point.
It consists of an inner tracking detector surrounded by a thin superconducting solenoid, electromagnetic and hadronic calorimeters,
and a muon spectrometer incorporating three large superconducting toroidal magnets.
 
The inner-detector system (ID) is immersed in a \SI{2}{\tesla} axial magnetic field
and provides charged-particle tracking in the range \(|\eta| < 2.5\).
The high-granularity silicon pixel detector covers the vertex region,
and is composed of four layers including
the insertable B-layer~\cite{ATLAS-TDR-19,PIX-2018-001}.
It is followed by the silicon microstrip tracker, which usually provides four two-dimensional measurement points per track.
These silicon detectors are complemented by the transition radiation tracker (TRT),
which enables radially extended track reconstruction up to \(|\eta| = 2.0\).
 
The calorimeter system covers the pseudorapidity range \(|\eta| < 4.9\).
Within the region \(|\eta|< 3.2\), electromagnetic calorimetry is provided by barrel and
endcap high-granularity lead/liquid-argon (LAr) electromagnetic calorimeters,
with an additional thin LAr presampler covering \(|\eta| < 1.8\)
to correct for energy loss in material upstream of the calorimeters.
The hadronic calorimeters have three sampling layers longitudinal in
shower depth in \(|\eta| < 1.7\) and four sampling layers in \(1.5 < |\eta| < 3.2\),
with a slight overlap in \(\eta\).
The solid angle coverage is completed with forward copper/LAr and tungsten/LAr calorimeter modules (FCal)
optimised for electromagnetic and hadronic measurements respectively.
 
The muon spectrometer (MS) comprises separate trigger and
high-precision tracking chambers measuring the deflection of muons in a magnetic field generated by the superconducting air-core toroids.
The field integral of the toroids ranges between \num{2.0} and \SI{6.0}{\tesla\metre}
across most of the detector.
A set of precision chambers covers the region \(|\eta| < 2.7\) with three layers of monitored drift tubes,
complemented by cathode strip chambers in the forward region, where the background is highest.
The muon trigger system covers the range \(|\eta| < 2.4\) with resistive plate chambers in the barrel, and thin gap chambers in the endcap regions.
 
The zero-degree calorimeters (ZDCs) are located symmetrically at $z = \pm 140$~m and cover $|\eta| > 8.3$ during the \pbpb\ data-taking period.
The ZDCs use tungsten plates as absorbers and quartz rods sandwiched between the tungsten plates as the active medium.
In Pb+Pb collisions the ZDCs primarily measure spectator neutrons,
which are neutrons that do not interact hadronically when the incident nuclei collide.
A ZDC coincidence trigger is implemented by requiring the pulse height from each
ZDC to be above a threshold set to accept the energy of a single neutron.
 
A two-level trigger system is used to select interesting events~\cite{ATLAS:2020esi}.
The first-level trigger is implemented in hardware and uses a subset of detector information,
including ZDC coincidences in \pbpb\ collisions,
to reduce the event rate to a design value of at most 100~kHz.
This is followed by a software-based high-level trigger which reduces the event rate to several kHz.
An extensive software suite~\cite{ATL-SOFT-PUB-2021-001} is used in the reconstruction and analysis of real
and simulated data, in detector operations, and in the trigger and data acquisition systems of the experiment.

% End of text imported from the .//chapters/detector.tex input file

\section{Data selection and Monte Carlo samples}
\label{sec:selection}

% The next lines are included from the .//chapters/data_selection.tex input file

All events are required to have at least one reconstructed
vertex, to have been collected during stable beam conditions,
and to satisfy detector and data-quality requirements~\cite{ATLAS:2019fst}.

These events were selected by two sets of triggers: single-jet triggers,
and muon--jet triggers requiring a muon spatially matched to a jet at the
trigger level~\cite{ATLAS:2021piz}. In both \pbpb\ and \pp\ collisions,
the muon--jet triggers require a muon with $\pt>$~4~\GeV\
matched to an \RFour\ jet with various \pt\ thresholds. The spatial matching
of the jet to the muon is made within a cone of size
$\Delta R =0.5$ around the jet axis; this matching is fully efficient.
All events are in a kinematic range where the jet trigger is fully efficient.

The overlap area of two colliding nuclei in Pb+Pb collisions is characterized by the total transverse energy
deposited in the FCal~\cite{HION-2016-06}. This analysis uses three centrality
intervals which are defined according to
successive percentiles of the $\sum E_\textrm{T}^\textrm{FCal}$ distribution obtained in MB
collisions.
The centrality regions used in this analysis, going from the most central
(largest $\sum E_\textrm{T}^\textrm{FCal}$) collisons to the peripheral (lowest $\sum E_\textrm{T}^\textrm{FCal}$)
collisions are 0--20\%, 20--50\%, and 50--80\%.
The values of  $\langle\TAA\rangle$
and the number of binary nucleon--nucleon collisions, $\langle N_\textrm{coll}\rangle$, for
each centrality interval are evaluated by a
Monte Carlo (MC) Glauber model analysis~\cite{Loizides:2014vua,Miller:2007ri} of the $\sum E_\textrm{T}^\textrm{FCal}$ distribution,
and are shown in Table~\ref{tab:TAANcoll}.
 
\begin{table}[htbp]
\centering
\caption{The $\langle\TAA\rangle$ and $\langle N_\textrm{coll}\rangle$ values and uncertainties in each centrality bin. These are the results from the Glauber modelling of the summed transverse energy in the forward calorimeters, $\sum E_\textrm{T}^\textrm{FCal}$.}
\vspace{0.2cm}
\begin{tabular}{ccc}
\toprule
Centrality [\%] &  $\langle\TAA\rangle$ $[\textrm{mb}^{-1}]$ & $\langle N_\textrm{coll}\rangle$ \\
\midrule
~~0--20         & 18.84$\pm$0.18~~     & 1319$\pm$90~~ \\
20--50       & 5.41$\pm$0.15       & 378$\pm$23\\
50--80       & 0.69$\pm$0.05       & 48$\pm$4\\
\bottomrule
\end{tabular}
\label{tab:TAANcoll}
\end{table}

A small fraction of the triggered \pbpb\ events $(< 0.5\%)$ contain multiple collisions, known as `pile-up'.
The expected
anti-correlation between $\sum E_\textrm{T}^\textrm{FCal}$
and the number of neutrons detected in the ZDCs is used to reject these pile-up events.
The pile-up contribution is not rejected in \pp\ collisions.

This analysis uses several MC samples to evaluate the performance of the detector
and the analysis procedure, and to correct the measured distributions for detector effects.
All MC samples were produced with the full ATLAS detector simulation based on the
$\GEANT4$ toolkit~\cite{Agostinelli:2002hh,Aad:2010ah}.
Dijet samples were generated with \pythiaeight~\cite{Sjostrand:2014zea}, using the parameter values of the
A14 tune~\cite{TheATLAScollaboration:2014rfk} with the \textsc{NNPDF23lo} set of parton
distribution functions (PDFs)~\cite{Ball:2012cx}.
These were generated both with no additional requirements and with the
requirement of a muon with $\pt >$~3~\GeV\
at the generator level (such as from heavy-flavour hadron decays).
The \EVTGEN  package~\cite{Lange:2001uf} was used to fully simulate \bhadron\ decays.
 
In \pp\ collisions,
pile-up from additional interactions in the same bunch crossing was generated by \pythiaeight,
using the parameter values of the A3 tune~\cite{ATLAS:2016puo}  with the \textsc{NNPDF23lo} PDFs;
the distribution of the number of extra collisions was matched to that of data.
In \pbpb\ collisions the simulated events were overlaid with events from a dedicated sample of \pbpb\ data events.
This sample was recorded with a combination of MB triggers and total energy triggers to
increase the number of events from central collisions. This `data overlay' sample was reweighted on an event-by-event
basis to obtain the same centrality distribution as in the jet-triggered data sample.
 
For muon reconstruction studies, prompt $(\pp\rightarrow J/\psi\rightarrow \mu\mu)$ and
non-prompt $(\pp\rightarrow bb\rightarrow J/\psi\rightarrow \mu\mu)$ samples of $J/\psi$
events
were produced with \pythiaeight\ and corrected for electromagnetic radiation with \textsc{Photos}~\cite{BARBERIO1994291}.
The A14 tune and CTEQ6L1 PDFs~\cite{Pumplin:2002vw} were used.
Generator-level \pp\ \herwig~\cite{Bahr:2008pv} events,
using the UEEE5 tune~\cite{Gieseke:2012ft} and the
CTEQ6L1 PDFs~\cite{Pumplin:2002vw}, were used to study systematic uncertainties.

% End of text imported from the .//chapters/data_selection.tex input file

\section{Analysis}
\label{sec:analysis}

% The next lines are included from the .//chapters/analysis.tex input file
\subsection{Jet reconstruction}
 
All jets are reconstructed using  procedures which
follow those used by ATLAS for previous jet measurements in \PbPb\ collisions~\cite{Aaboud:2018twu,ATLAS:2021ktw}.
Jets are reconstructed using
the \antikt\ algorithm~\cite{Cacciari:2008gp} implemented in the FastJet software package~\cite{Cacciari:2011ma}. In both \pp\ and \pbpb\ collisions, jets with
$R=0.2$ and \RFour\ are formed by
clustering calorimetric towers of spatial size $\Delta\eta \times\Delta\phi=0.1\times\pi/32$. The energies in the towers are obtained
by summing the energies of calorimeter cells at the electromagnetic energy scale~\cite{PERF-2011-03} within the tower boundaries.
In \pbpb\ collisions, a background
subtraction procedure is applied to estimate, within each event, the underlying event (UE) average transverse energy density,
$\rho(\eta,\phi)$, where the $\phi$ dependence is due to global azimuthal correlations in the particle production from
hydrodynamic flow~\cite{HION-2016-06}. The modulation accounts for the contribution to the UE of the second-, third-, and fourth-order azimuthal anisotropy harmonics
characterized by values of flow coefficients $v_n^\mathrm{UE}$~\cite{HION-2016-06}.
Additionally, the
UE is also corrected for $\eta$- and $\phi$-dependent non-uniformities of the detector response
by correction factors derived in MB \PbPb\ data.
In \pp\ collisions, the same background subtraction procedure is applied without the $\phi$-dependent
modulation and without the correction for $\eta$- and $\phi$-dependent non-uniformities
to remove the pile-up contribution to the jet.
 
An iterative procedure is used to remove the impact of jets on the estimated $\rho$ and $v_n^\mathrm{UE}$
values.  The first estimate of the average transverse energy density of the UE, $\rho(\eta)$, is evaluated in 0.1 intervals of $\eta$, excluding towers
within $\Delta R =0.4$ of `seed' jets. In the first subtraction step, the seeds are defined to be an union of $R = 0.2$ jets and $R=0.4$ track-jets. Track-jets are
reconstructed by applying the \antikt\ algorithm with $R = 0.4$ to charged particles with $\pT > 4$~\GeV. The $R=0.2$ jets must pass a requirement on the
minimum value of the tower $E_{\mathrm{T}}$  and on a ratio of maximum tower  $E_{\mathrm{T}}$ to
average tower  $E_{\mathrm{T}}$,
while the track-jets are required to have $\pT > 7$~\GeV. The background is then subtracted from each tower
constituent and the jet kinematics are recalculated. After the first iteration, the $\rho$ and $v_n$ values are updated by excluding from the UE determination
the regions within $\Delta{R} = 0.4$ of both the track-jets and the newly reconstructed $R=0.2$ jets with $\pt > 25$~\GeV\ (8~\GeV) in \pbpb\ (\pp) collisions. The updated
$\rho$ and $v_n^\mathrm{UE}$ values are used to update the jet kinematic properties in the second iteration.
 
Jet $\eta$- and $\pT$-dependent correction factors derived in simulations are applied to the measured jet
energy to correct for the calorimeter energy response~\cite{ATLAS:2014hvo}. An additional correction based on in situ studies of jets recoiling
against photons, \Zboson\ bosons, and jets in other regions of the calorimeter is applied~\cite{HION-2018-06}. This calibration is followed by a
cross-calibration which relates the jet energy scale of jets reconstructed by the procedure outlined in this section to the jet energy scale
in 13~\TeV\ \pp\ collisions~\cite{ATLAS:2015twa}.

Jets are defined at the truth level in the MC sample before detector simulation by applying the \antikt\
algorithm with the appropriate $R$ value
to stable particles with a proper lifetime greater than 30~ps, but excluding muons and neutrinos, which do not
leave significant energy deposits in the calorimeter.
The \dR\ between the truth jet and reconstructed jet is required to be $\dR < 0.15$ (0.30) for \RTwo\ (\RFour) jets.
The  corrections for muons and neutrinos from semileptonic decays are discussed below.

\subsection{Muon reconstruction}
Muon candidates in both \pbpb\ and \pp\ collisions
are formed by combining charged-particle tracks reconstructed
in the ID and the MS that pass the `tight' selection requirements detailed in Ref.~\cite{ATLAS:2016lqx},
except the requirement on the number of TRT hits.
 
Muons are selected with $\ptmu  > 4$~\GeV\ and $|\eta^{\mu}|<2.4$ requirements.
If the muon is required in the  trigger,
the data is corrected to account for the muon trigger inefficiency, and the reconstructed muon must be within
$\Delta R =0.01$ of the trigger muon object.
The muon selection is the same in both \pp\ collisions and \pbpb\ collisions.
Muons and jets are judged to be associated if $\dR(\textrm{jet},\mu)$ is less than the jet
radius, $R$. If more
than one muon passes this selection, the muon with the largest momentum is used.
 
The muon trigger efficiency with respect to reconstructed muons
is estimated from \pp data by using the tag-and-probe method~\cite{Aad:2020uyd,ATLAS:2021xtw}
in bins of \ptmu\ and $\eta^{\mu}$. A small centrality dependence of the muon trigger efficiency was observed in Pb+Pb data;
this is corrected by an extra factor, which is the
Pb+Pb to \pp data-driven efficiency ratio, applied to Pb+Pb data as a function of the centrality.
In \pp collisions, the trigger efficiency plateaus at 78\% for $\ptmu > 6$~\GeV\
and $|\eta^{\mu}|<1.05$, and at 90\% for $\ptmu > 9$~\GeV\ and $|\eta^{\mu}|>1.05$.
In \pbpb\ collisions,
the efficiency is lower than in \pp\ collisions by 9\% in the most central collisions;
the efficiency in peripheral \pbpb\ collisions is the same as in \pp\ collisions.

The muon reconstruction efficiency is estimated from simulated prompt and non-prompt $J/\psi\rightarrow\mu\mu$ events using the tag-and-probe method~\cite{ATLAS:2016lqx,ATLAS:2021xtw},
in fine bins of \ptmu\ and $\eta^{\mu}$. Mis-modelling of the reconstruction performance in simulation, quantified by the ratio of measured
efficiencies in \pp data and the simulation, is accounted for by applying a multiplicative
correction which changes the efficiency by less than 5\%.
For $\ptmu > 10$~\GeV\ in \pp\ collisions, the muon reconstruction efficiency plateaus at 90\% for $|\eta^{\mu}|<1.05$ and at 95\% for $|\eta^{\mu}|>1.05$.
No difference between the muon reconstruction efficiencies in \pp\ and \pbpb\ collisions was observed and the same
values were used in both collision systems.

\subsection{\bjet yield reconstruction}
\subsubsection{Templates and fitting}
 
At the generator level, the jet flavour is defined by matching jets to hadrons with $\pt > 5$~\GeV.
The jet is considered a \bjet\ if a \bhadron\ is found within $\Delta R =0.3$ of the jet axis;
otherwise, if a \chadron\ is found within the same distance the jet is labelled as a \cjet.
All other jets are considered to be \ljets.
 
Since the muons in the momentum range of interest generally
do not stop in the calorimeter, their momentum is not included in
the calorimetric jet \pt.  In order to better characterize the jets, the \jetmu\ scale
is defined as
the \pt of the sum of the muon and jet four-vectors.
The \ptrel\ distributions in the data and MC samples are constructed for selections in
this \jetmu\ object at the reconstruction level, \ptrecojetmu.
In order to form a \jetmu\ pair, jets are required to have a
minimum calorimetric \ptrecojet\ of 40~\GeV\ (58~\GeV)
in \pp\ (\pbpb) collisions.
 
In semileptonic \bhadron\ decays, the \pt of the lepton relative
to the \jetmu-axis (see Eqs.~(\ref{eq:jetmuaxis}) and~(\ref{eq:ptrel})),
\ptrel, is used to distinguish between $b$-, $c$-, and \ljets~\cite{ATLAS:2011ac}.
Due to the large mass of \bhadrons\ relative to hadrons containing charm or light quarks,
their decay products are more energetic.
Consequently, muons originating from \bhadron\ decays have a harder \ptrel\
spectrum than muons in \cjets\ and \ljets.
 
The templates distributions of \ptrel\
for $b$- and \cjets\  are extracted from the muon-filtered MC samples.
The \bhadron\ mixture is taken from Tevatron measurements~\cite{Amhis:2014hma},
which are consistent with
those from the LHCb Collaboration~\cite{LHCb:2019fns}.
The \chadron\ mixture is taken from default \pythiaeight\ simulation.
The \ljet\ template is obtained from track-jet pairs in the data inclusive-jet sample
as in Refs.~\cite{CMS:2011xhf,ATLAS:2011ac} (in the same centrality class as the \jetmu\ pair for \pbpb\ collisions).
The tracks are selected with the standard selection,
with no additional requirements on the distance of closest approach to the vertex.
The \pt\ spectra of the tracks
are reweighted to reproduce the muon \pt spectrum from \ljets\ from the inclusive dijet MC sample.
 
In \pp\ collisions, these three templates are inputs to a template fit of the
data \ptrel\ spectrum
to obtain the fractions of $b$-, $c$-, and \ljets.
A binned maximum-likelihood fit is
performed using the RooFit framework~\cite{Verkerke:2003ir}.
The model for the fit is
defined as a normalized sum of three templates:
\ljets\ (\Fl(\ptrel)), \cjets\ (\Fc(\ptrel)), and \bjets (\Fb(\ptrel)). The \pp fit model,
$M^{\pp}$, is structured as
\begin{equation*}
M^{\pp}_{i}\!\left(\ptrel\right) = \fb \Fb\!\left(\ptrel\right) +
\left(1-\fb\right)\left[ \fc \Fc\!\left(\ptrel\right) +
\left(1-\fc\right)\Fl\!\left(\ptrel\right) \right]
\label{eq:ppmodel}
\end{equation*}
where \fb\ is the fraction of \bjets\ in the total sample,
\fc\ is the fraction of \cjets\ in the non-\bjet\ sample, and the index $i$ denotes the \pt bin.
The fit model has two free parameters, \fb\ and \fc\ for each \ptrecojetmu\ bin.
 
In \pbpb\ collisions the model has an additional term
which accounts for combinatoric muon--jet pairs.
These are muon--jet pairs which pass the $\Delta R$ matching requirements, but the
muon and jet  do not come from the same hard scattering.
These are the most common in central \pbpb\ collisions, where the multiplicity
of jets and muons is the highest.  In \pp\ collisions, this
effect is negligibly small.  For the combinatoric pairs, both the shape, \Fbkg,
and the amplitude, \fbkg, are determined by event mixing in the \pbpb\ data.
This procedure uses a muon from one event and a jet from another event;
all the analysis selections are applied, including the requirement on the $\Delta R$ separation between the muon
and jet.
Events which are mixed together  are required to have $\sum E_\textrm{T}^\textrm{FCal}$ values that agree within 0.05~\TeV\
and vertex $z$-positions that agree within 10~mm.
The nominal data sample and mixed-event sample are found to agree
well in the large $\Delta R$ region where there is no correlated signal.
The \pbpb\ model, $M^\textrm{Pb+Pb}$, is structured as
\begin{equation*}
M^\textrm{Pb+Pb}_{i}\!\left(\ptrel\right) =
\left(1-\fbkg\right)\times M^{\pp}_{i}\!\left(\ptrel\right) + \fbkg\times \Fbkg\!\left(\ptrel\right).
\label{eq:pbpbmodel}
\end{equation*}
The background fraction is largest in central collisions and is always less than 1.5\%.
 
\begin{figure}[htbp]
\centering
\includegraphics[width=0.4\textwidth]{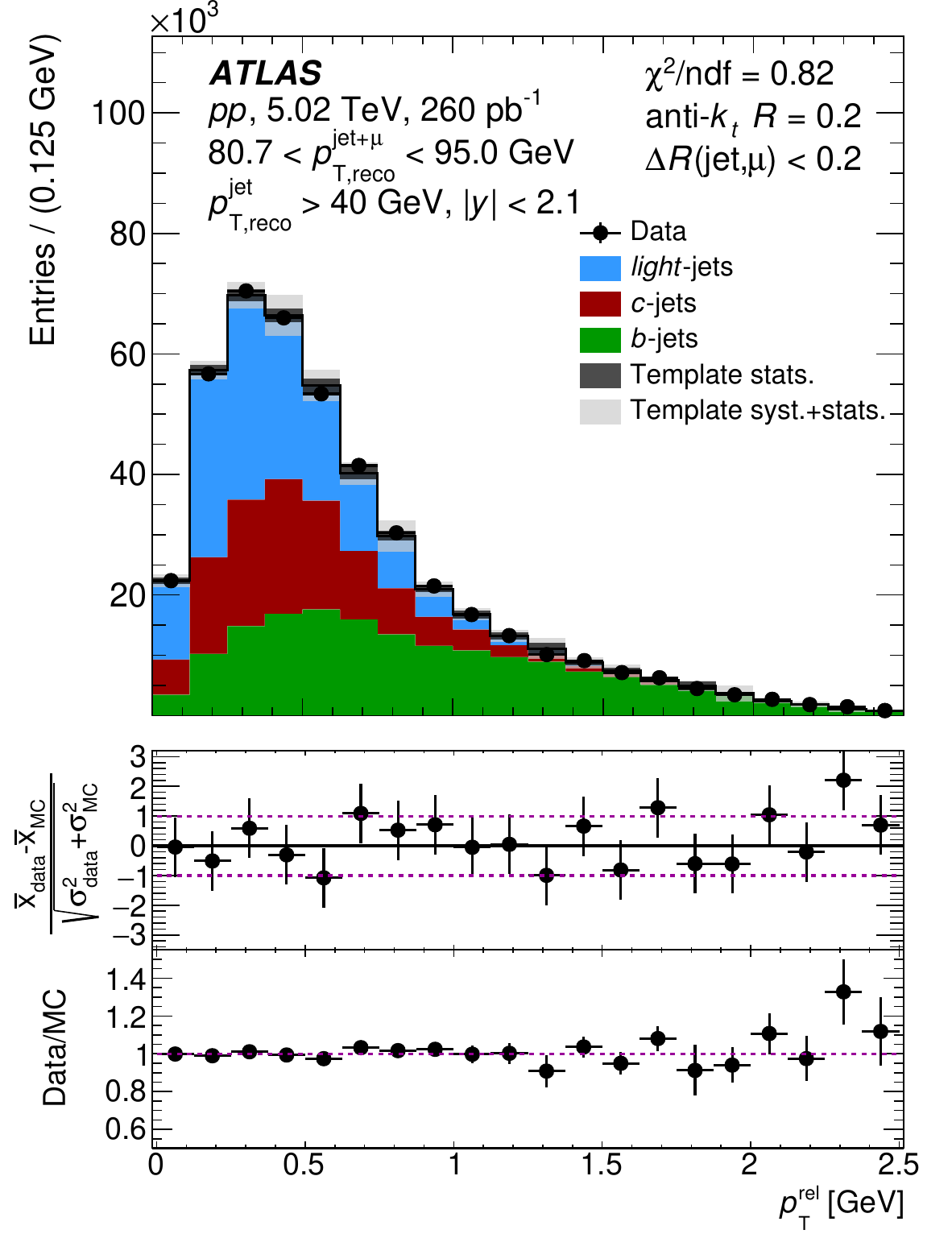}
\includegraphics[width=0.4\textwidth]{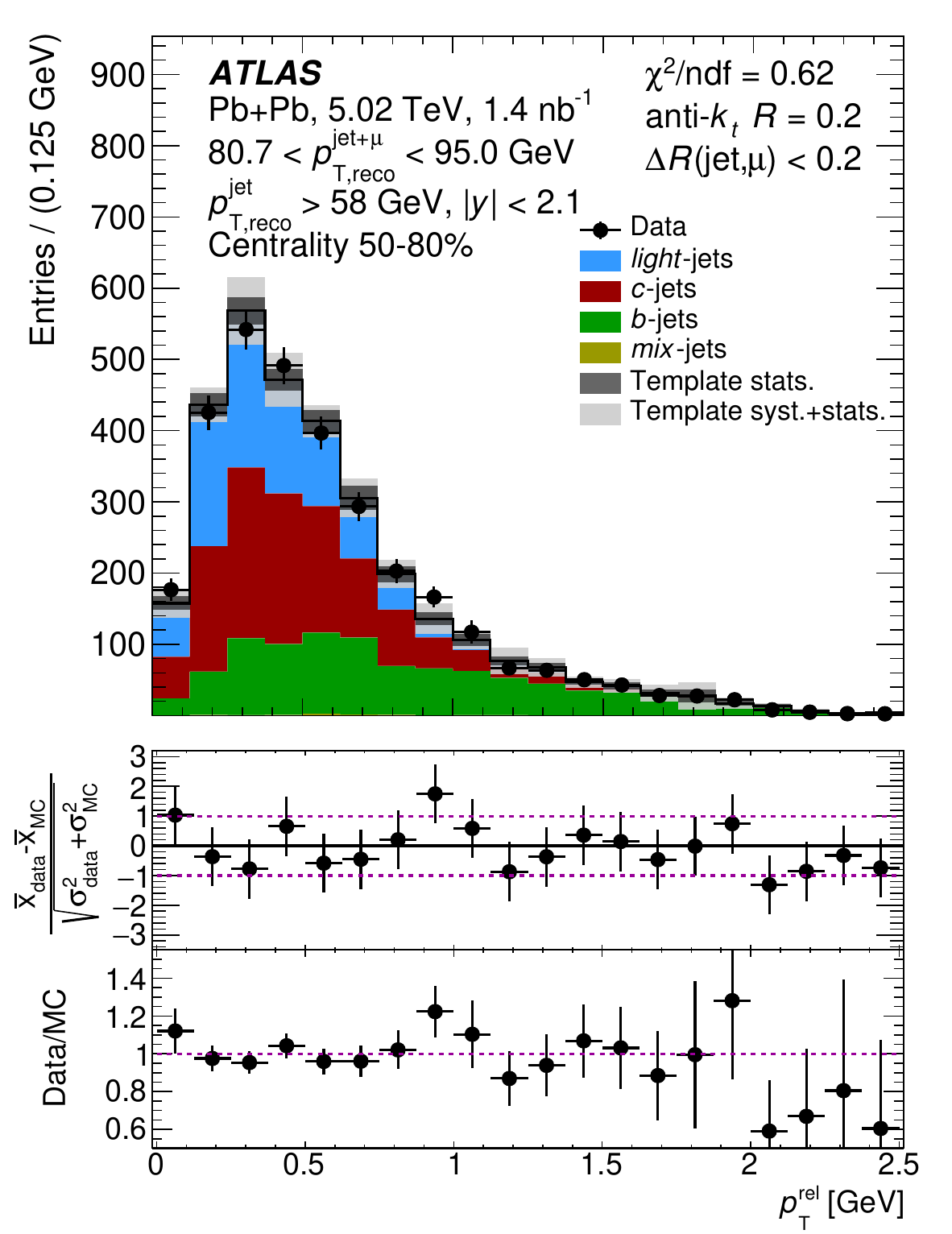}
\includegraphics[width=0.4\textwidth]{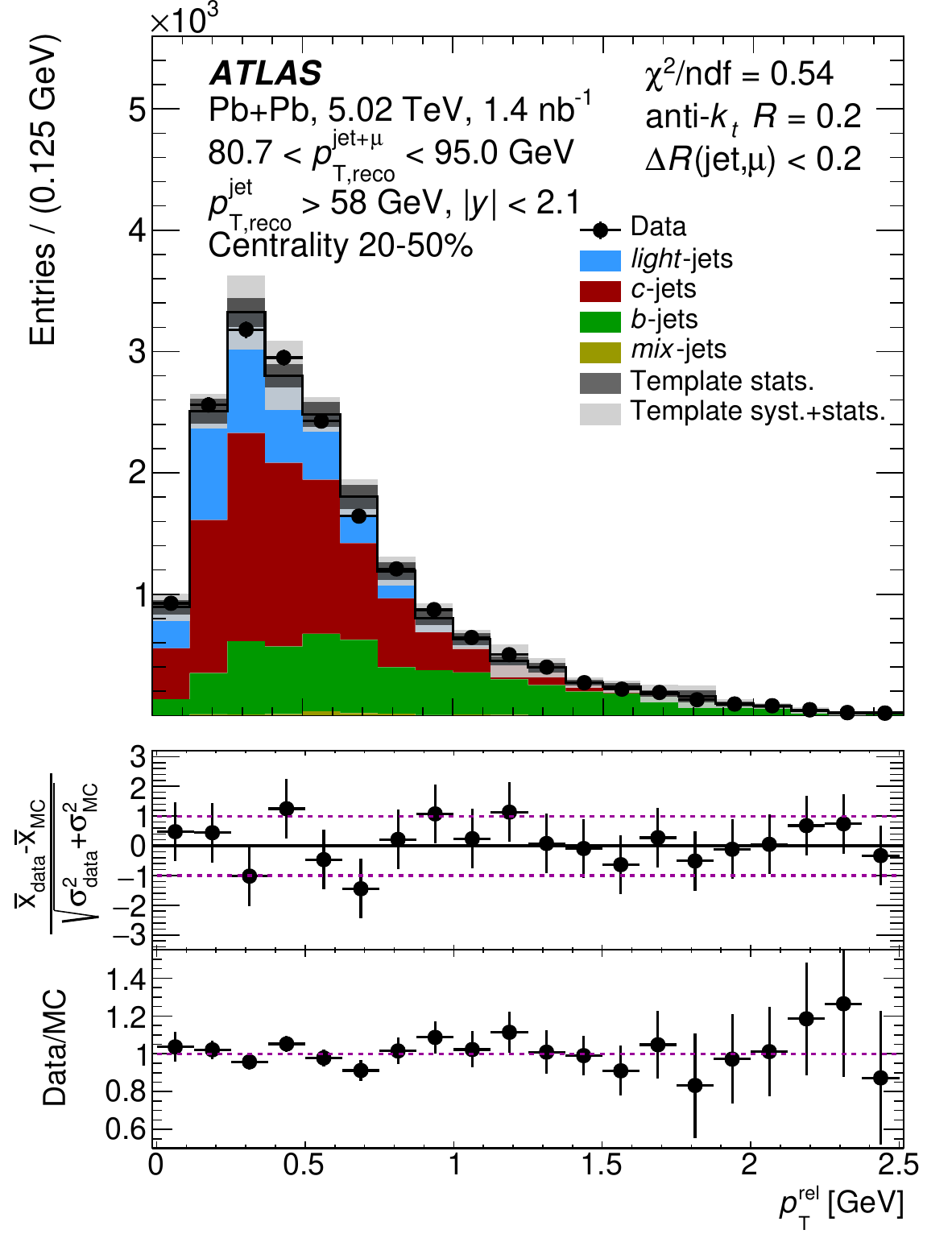}
\includegraphics[width=0.4\textwidth]{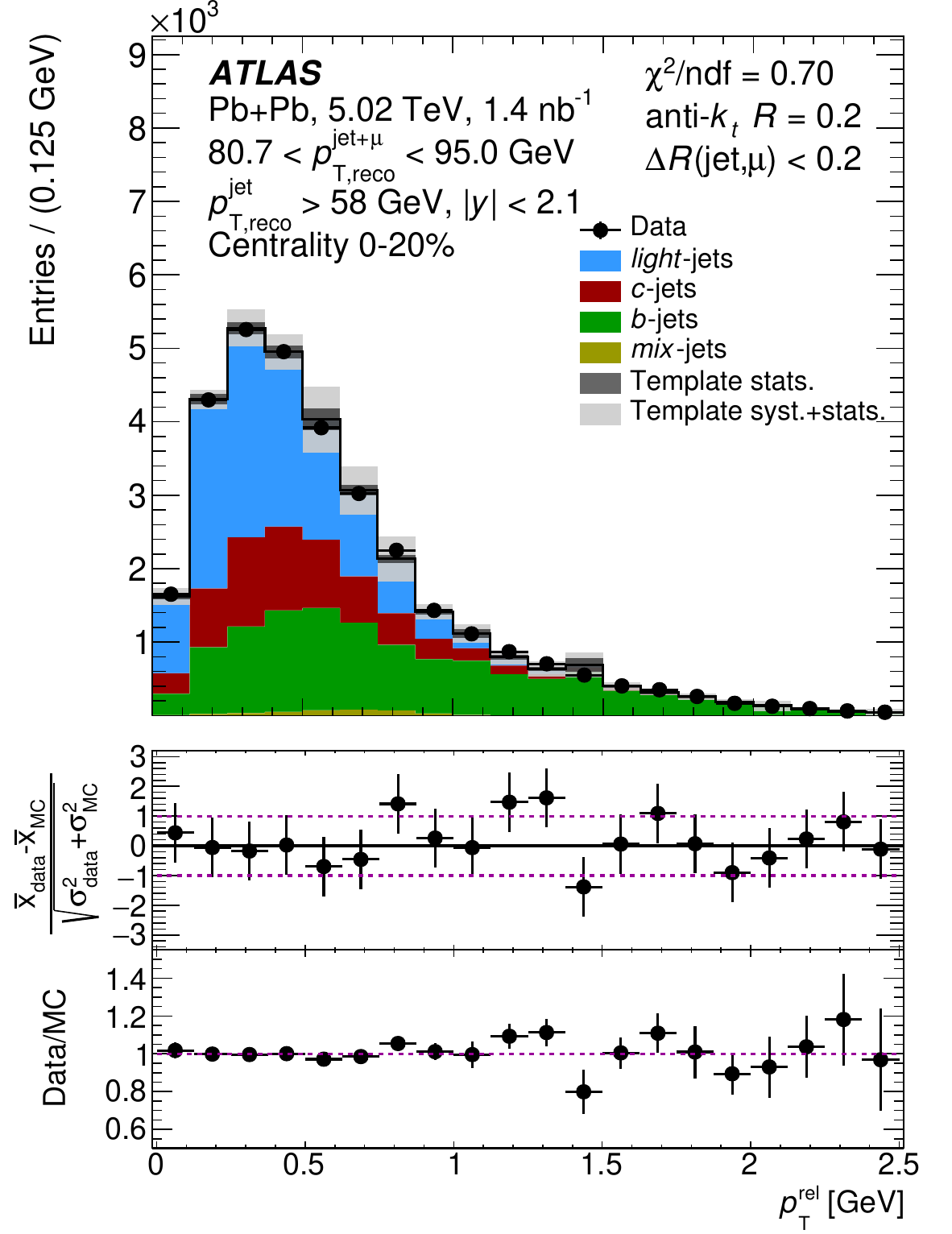}
\caption{The \ptrel\ distributions in \pp collisions (top left) and 50--80\% (top right), 20--50\% (bottom left),
and 0--20\% (bottom right) centrality \pbpb\ collisions.
The data are shown for
the range $80.7 < \ptrecojetmu < 95.0$~\GeV. The stacked histograms
show the fit results.
Middle and bottom panels of each plot show the pulls and data-to-fit ratios, respectively.
The error bars in the ratio plots show the statistical uncertainties of the data and MC
samples.}
\label{fig:fitslow}
\end{figure}

\begin{figure}[htbp]
\centering
\includegraphics[width=0.4\textwidth]{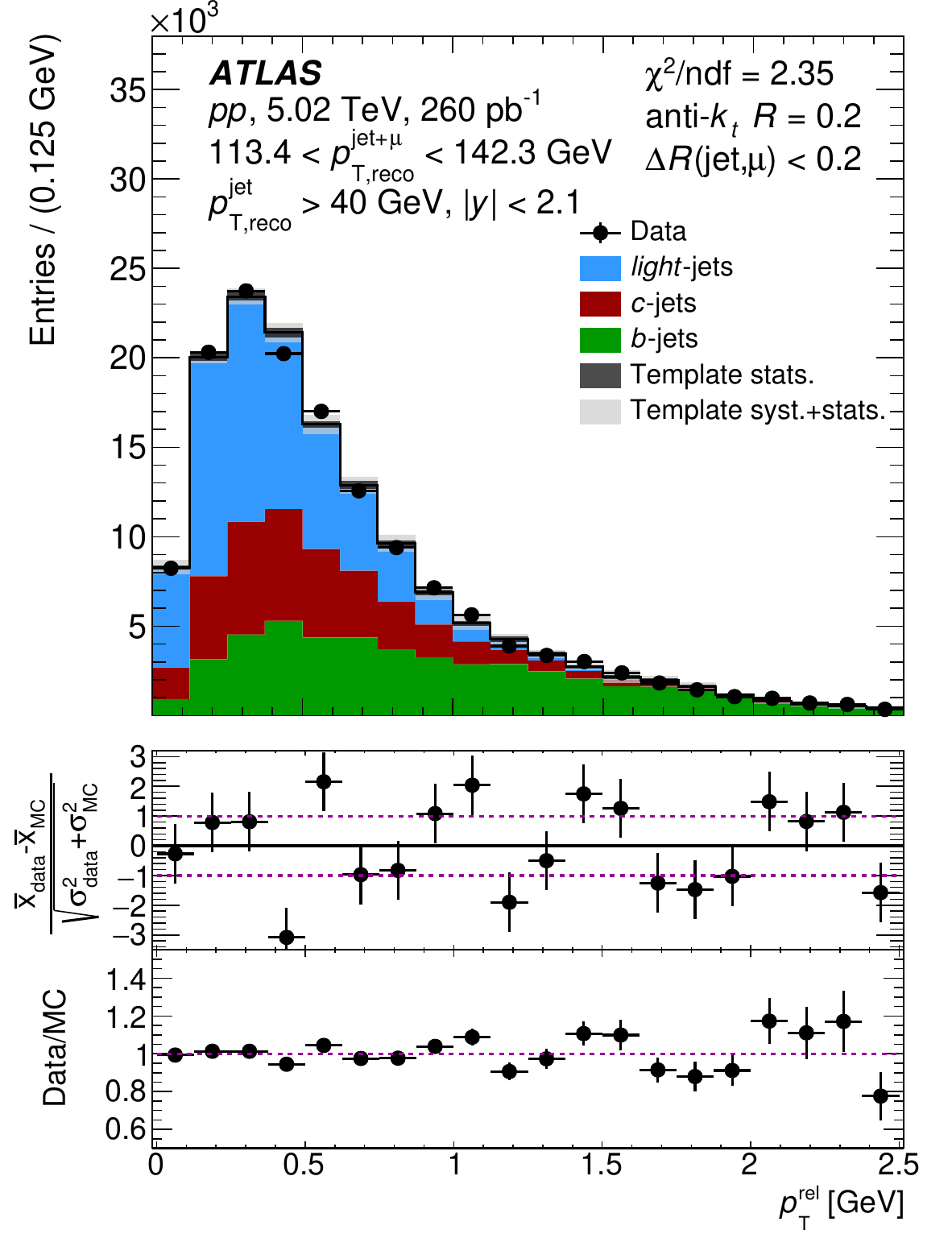}
\includegraphics[width=0.4\textwidth]{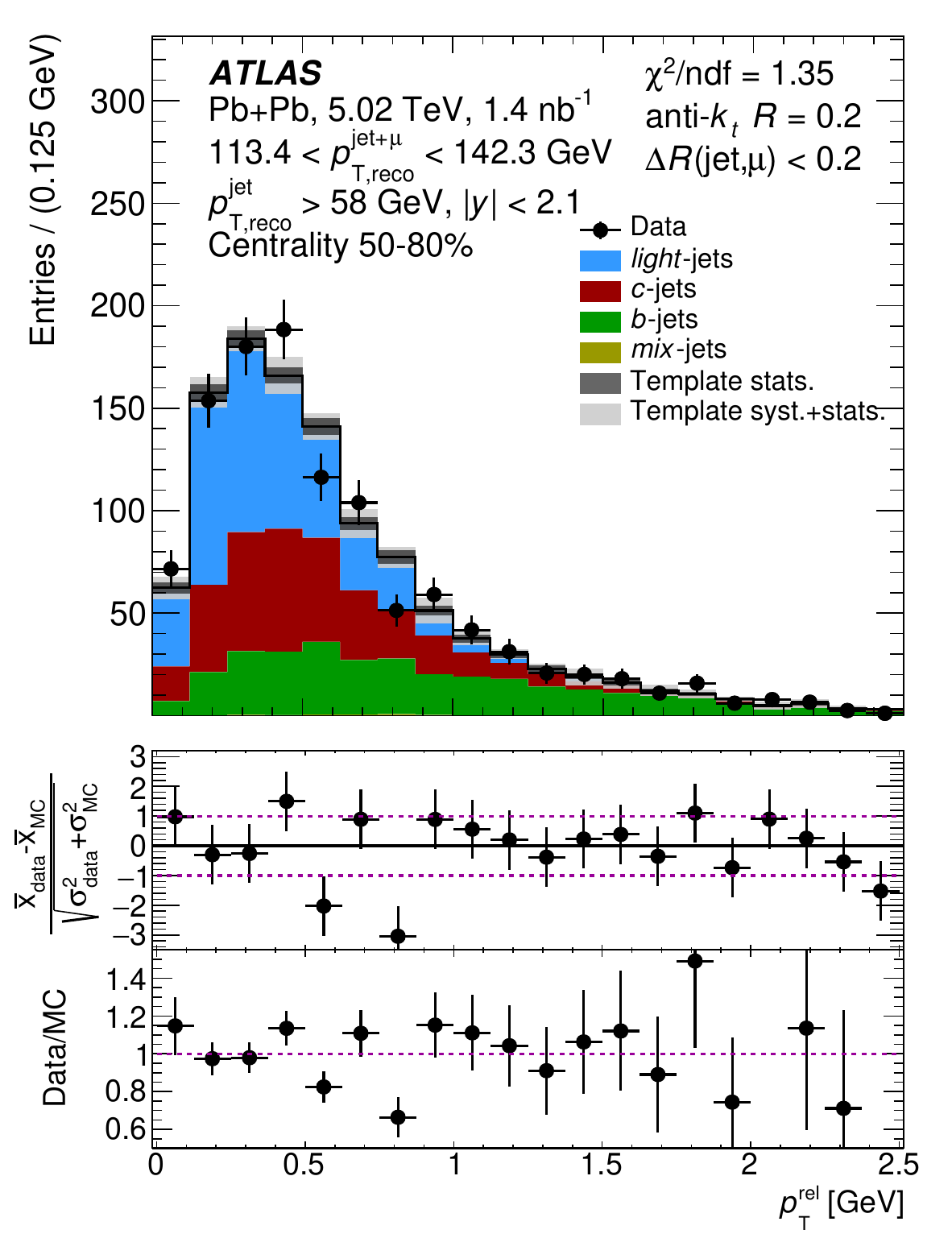}
\includegraphics[width=0.4\textwidth]{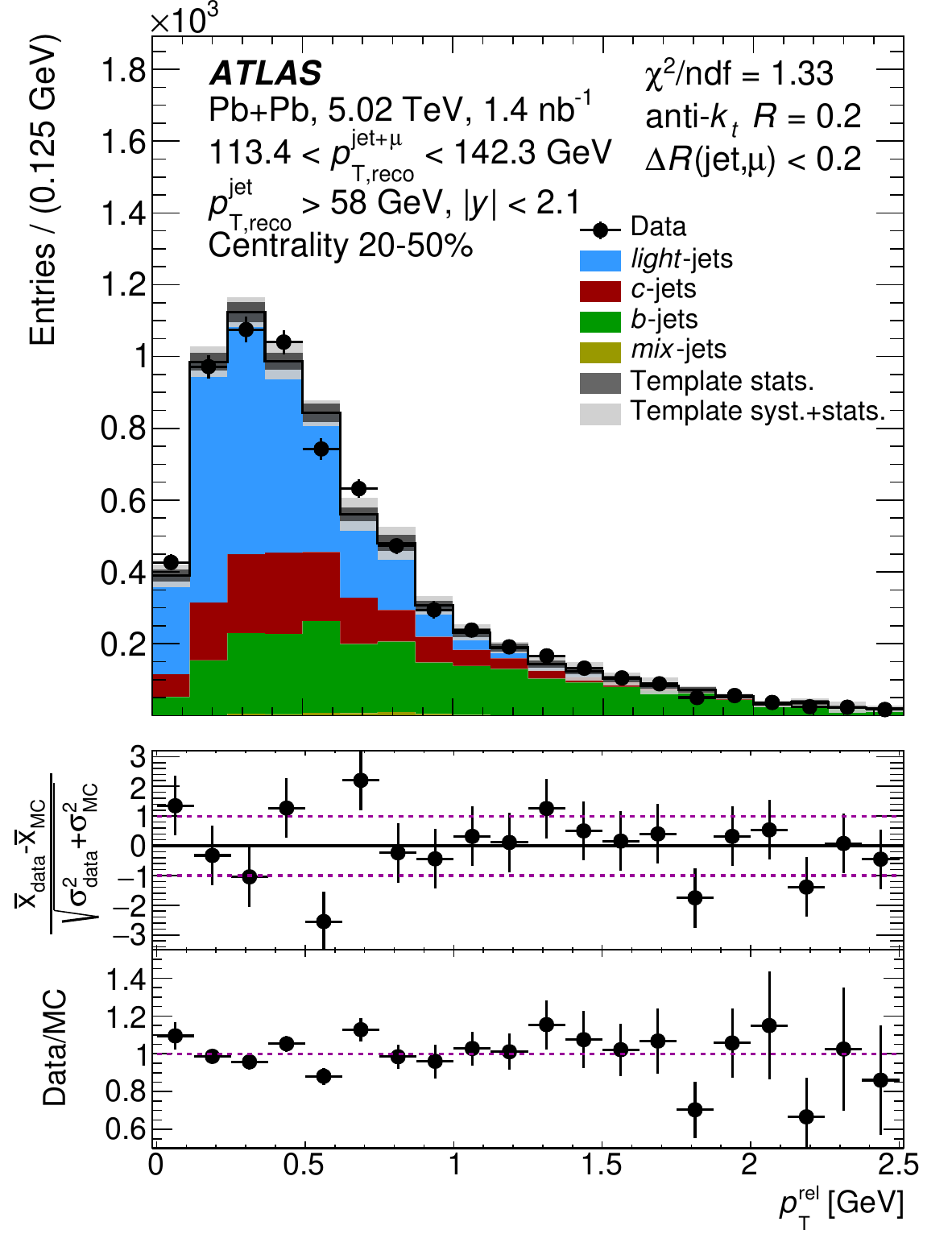}
\includegraphics[width=0.4\textwidth]{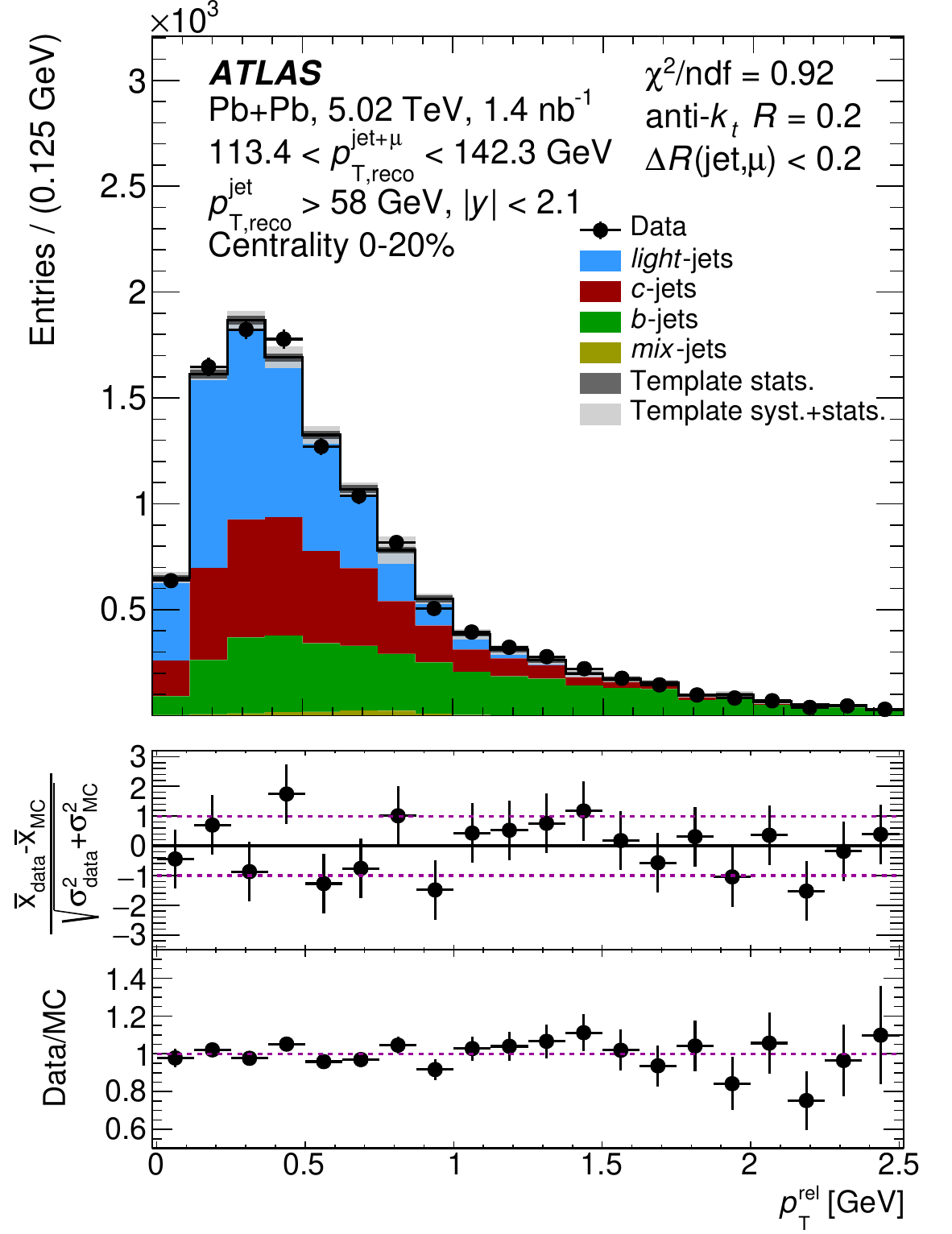}
\caption{The \ptrel\ distributions in \pp collisions (top left) and 50--80\% (top right), 20--50\% (bottom left),
and 0--20\% (bottom right) centrality \pbpb\ collisions.
The data are shown for
the range $113.4 < \ptrecojetmu < 142.3$~\GeV. The stacked histograms
show the fit results.
Middle and bottom panels of each plot show the pulls and data-to-fit ratios, respectively.
The error bars in the ratio plots show the statistical uncertainties of the data and
MC samples.}
\label{fig:fitshigh}
\end{figure}

The data \ptrel\ distributions are resampled 500 times and the mean and width of the range of
results are used to set the central value and uncertainties in the fit.
The statistical uncertainty of the templates is taken to be part of the statistical uncertainty
of the yield. Example fits for \pp and  \pbpb\ collisions are shown in Figures~\ref{fig:fitslow} and~\ref{fig:fitshigh},
for low and high \ptrecojetmu\ respectively.
 
The raw \bjet\ \pt\ spectrum
is constructed in each \pbpb\ centrality bin and in \pp\ collisions by taking
each \fb\ value and multiplying it by the number of jets in \ith\ bin in the total \ptrecojetmu\ spectrum.
\subsubsection{Comparing muon momentum distributions in the data and simulations}
 
\begin{figure}[htbp]
\centering
\includegraphics[width=0.49\textwidth]{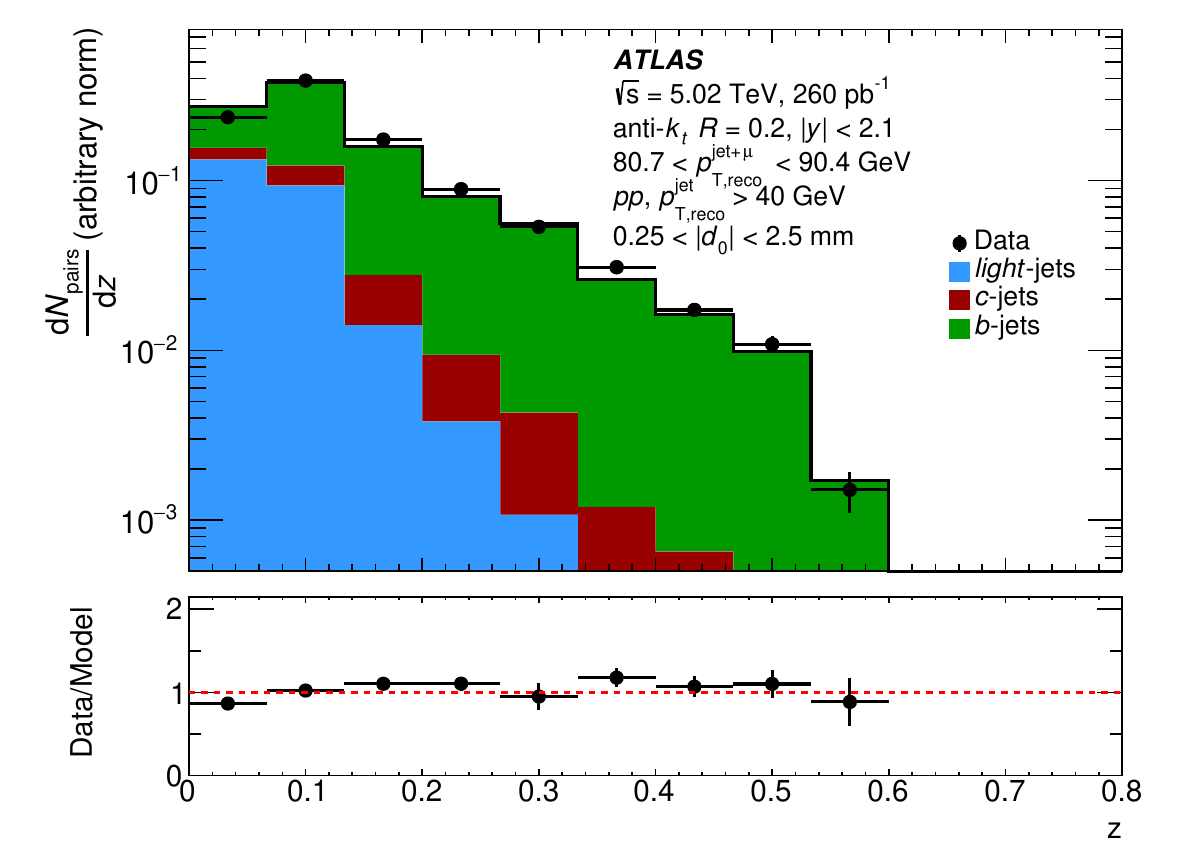}
\includegraphics[width=0.49\textwidth]{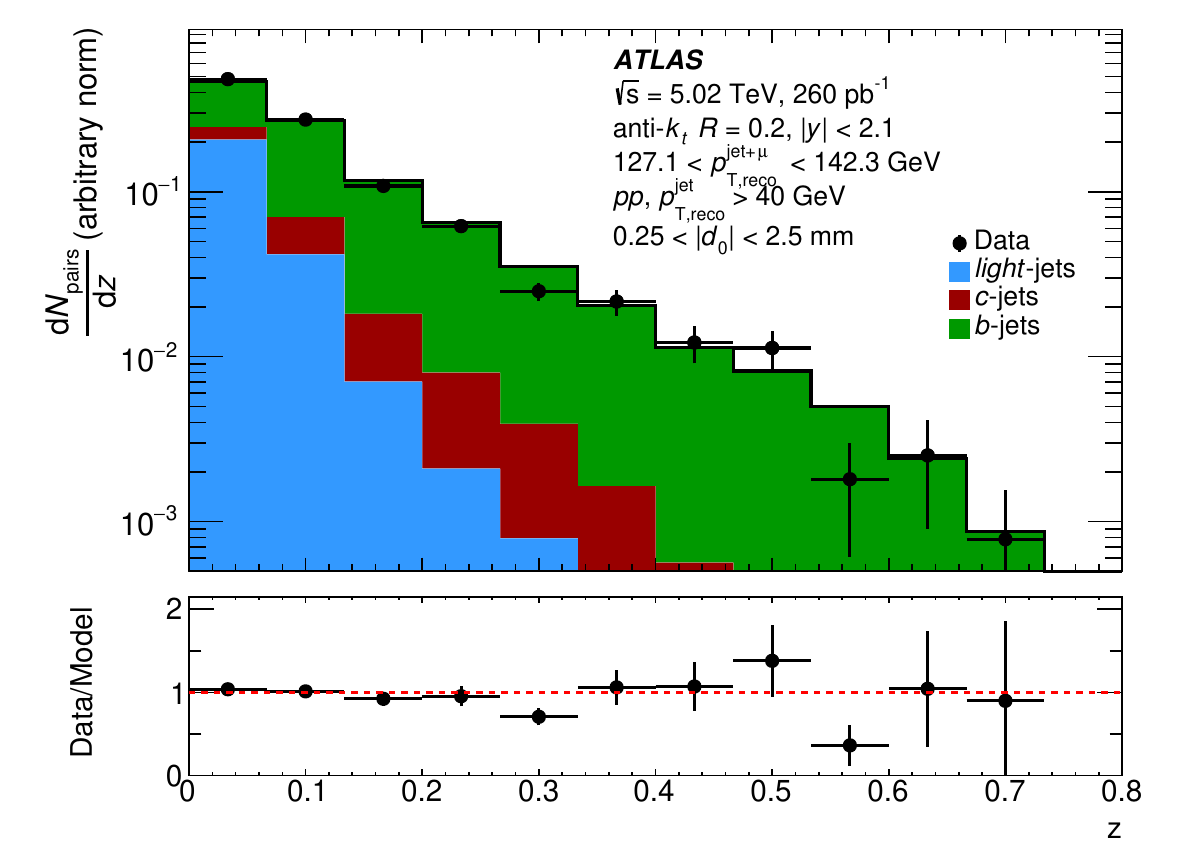}\\
\includegraphics[width=0.49\textwidth]{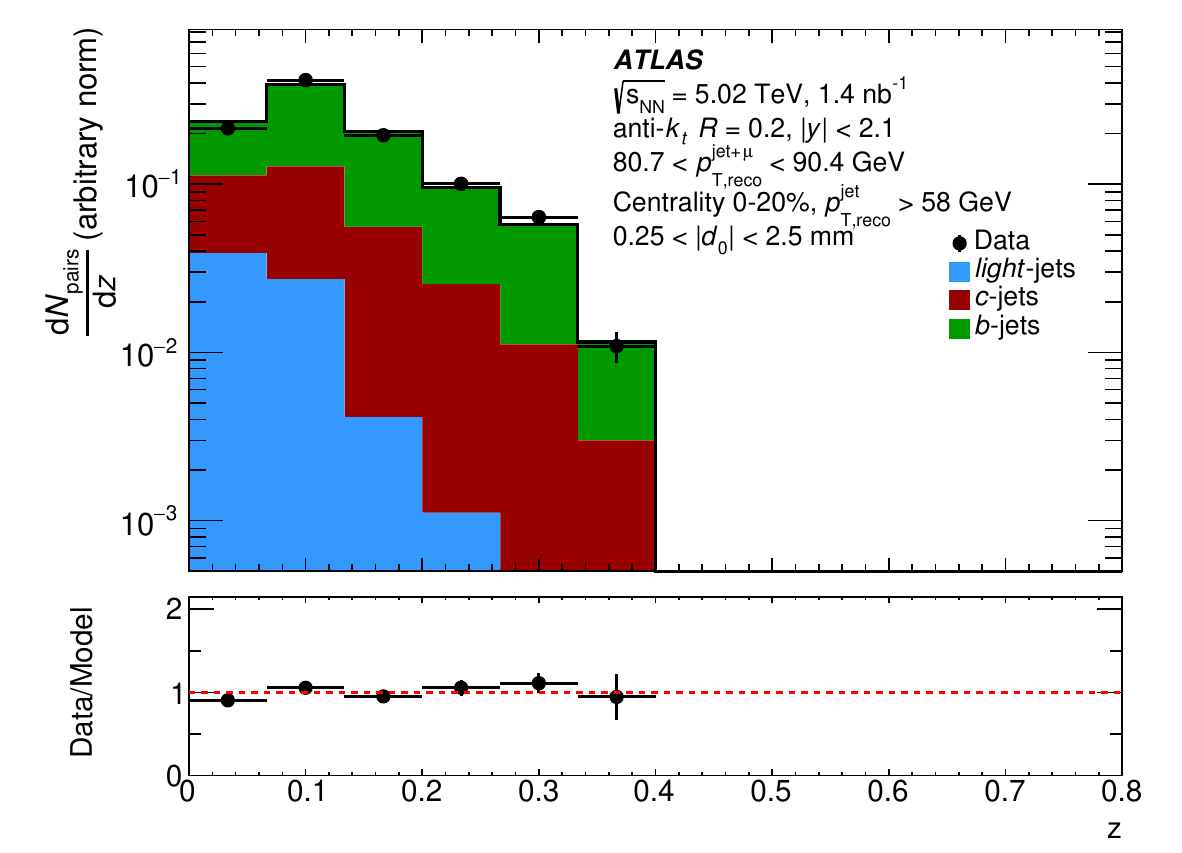}
\includegraphics[width=0.49\textwidth]{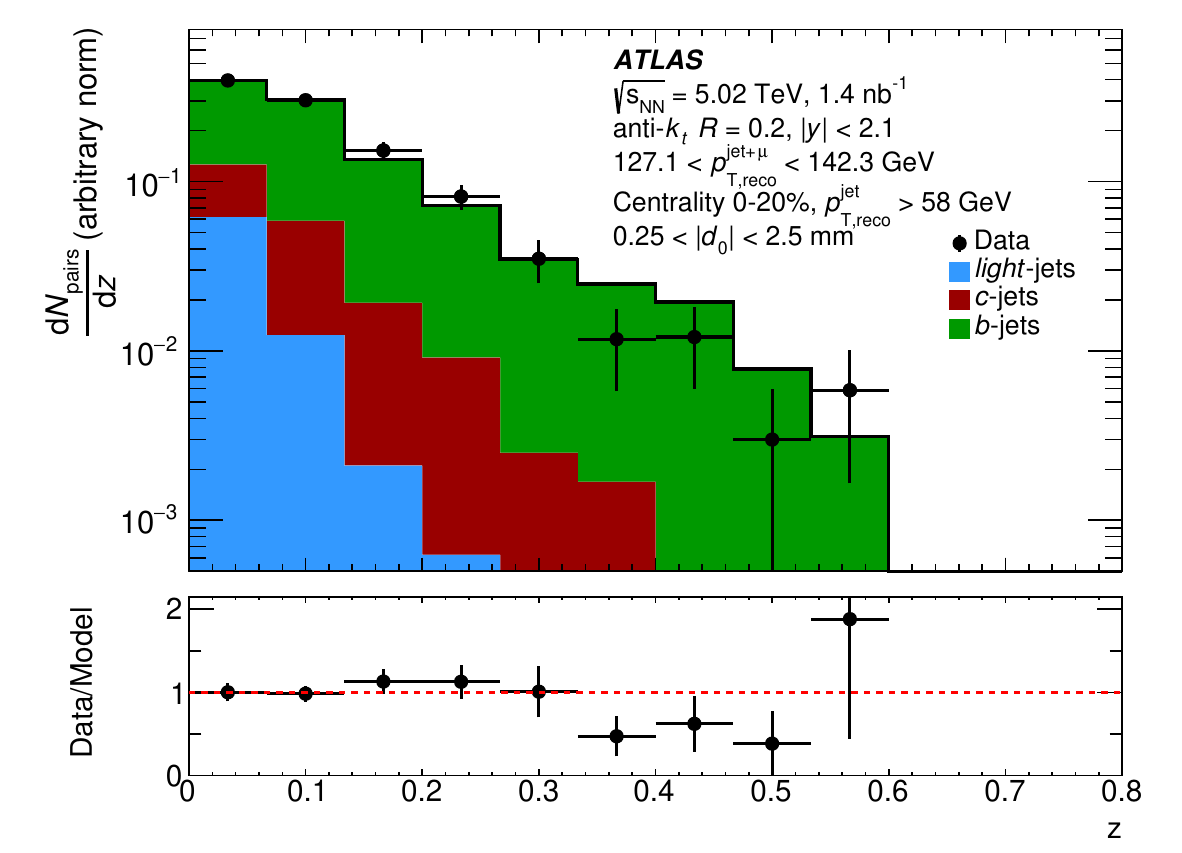}
\caption{The \z\ distributions for data and simulations (including the data-driven
\ljet\ template) for \pp\ collisions (top row) and 0--20\% centrality \pbpb\ collisions
(bottom row) for two \ptrecojetmu\ selections.  In each figure the top panel shows the distributions
themselves and the bottom panel shows the ratio of data to simulations.
The error bars in the ratio plots show the statistical uncertainties on the data and MC samples.}
\label{fig:longmuon}
\end{figure}

The \bjet\ fragmentation function in \pbpb\ collisions has not been
measured. Significant
modification to the \bhadron\ longitudinal momentum due to jet quenching could bias the \bjet\ template.
In order to ensure that the simulations reproduce the data sufficiently well, a check of
the longitudinal momentum distribution of the muon with respect to the \jetmu\ axis
is performed.  The fitting results as a function of centrality and \ptrecojetmu\ from the
previous subsection are used to weight the
\jetmu\ pairs from the MC samples (used for the charm and bottom templates) and those used in
constructing the data-driven \ljet\ template to have the same
flavour fractions as the data.  Muons which have a
transverse distance to the primary vertex, $d_0$, greater than 0.25~mm are selected;
this requirement further enhances the heavy-flavour fractions relative to the light jets.
For these \jetmu\ pairs,  the distribution of
\begin{equation*}
\z \equiv \ptmu \cos\!\left(\theta\right) / \ptrecojetmu\ ,
\end{equation*}
where $\theta$ is the angle between the muon and the \jetmu\ axis,  is constructed.
This combined distribution from the pairs in the charm-, bottom-, and \ljet\ templates is compared with the
data distribution for the same quantity.  Since the \ljet\ template is data-driven, it should
agree between the data and the template, and thus
the comparison between data and simulations is sensitive to
the degree to which the MC modelling of the charm and bottom templates
agrees with the data.
These distributions are shown in Figure~\ref{fig:longmuon} for \pp\ and
0--20\% centrality \pbpb\ collisions.  The data and the simulations
(including the data-driven \ljet\ template)
agree in both \pp\ and \pbpb\ collisions.
Thus, within the current uncertainties, the longitudinal momentum fraction
of the muon is unmodified in \pbpb\ collisions  and  the \ptrel\ distributions are
largely sensitive to the \bhadron\ decay kinematics.
 
\subsection{Corrections to the raw spectra}
The raw jet spectra are constructed from all the measured jets for the inclusive jets and from
the muon--jet pairs, scaled by the \fb\ values, for the \bjets after correction for the muon trigger efficiency.
The raw spectra are unfolded to account for bin migrations due to the finite jet energy resolution,
and any reconstruction inefficiency for the jets and muons (and muon and neutrino energies in the \bjet\ sample).
Both the \bjet\ and inclusive jet \pt\ spectra are unfolded using the one-dimensional (1D) Bayesian
unfolding~\cite{DAgostini:1994fjx} from the \textsc{RooUnfold} software package~\cite{Adye:2011gm}.
The \bjet\ response matrices are built from the muon-filtered MC samples
using truth-level \bjets\ (including the muon and
neutrino momenta)
that are matched to reconstructed jets in simulations
and have $|y_\textrm{truth}^{\bjet}|<2.1$.
The inclusive jet response matrices are built from the dijet MC samples using all
truth jets and not including any muon or neutrino momenta.
 
The response matrices are generated separately for \pp collisions and for each
centrality interval in \pbpb\ collisions.
To better represent the data, the response matrices are reweighted
along the truth-\ptjet\ axis by the reconstruction-level data to simulation ratio.
The number of iterations in the unfolding was chosen such that the result is stable
when changing the number of iterations while minimizing the amplification of statistical uncertainties. Four
iterations are used in all cases presented here.
The reconstructed \ptrecojetmu\ values are required to be larger than 72~\GeV\ in \pbpb\
collisions and 64~\GeV\ in \pp collisions for both \RTwo\ and \RFour\ jets. The inclusive-jet
minimum
reconstructed \pt is required to be larger than 72~\GeV\ for both \pbpb\ and \pp\ collisions.
Reconstructed jets below these thresholds that
match to truth-\pt values in the measurement region are
corrected via the unfolding procedure.
Fully unfolded results are presented starting from 80~\GeV\ for both categories of jets.
 
As described in Section~\ref{sec:selection}, the main MC sample used in the \bjet\ analysis required
the truth-muon \pt to be larger than 3~\GeV.
The missing part of the cross-section from \bjets\ with  muons below this threshold
is addressed through an acceptance correction based
on the \pythiaeight\ dijet MC samples
which is applied after unfolding. The size of the correction is
approximately 22\% (25\%)  at 80~\GeV\ and 19\% (23\%) at 250~\GeV\ for \RTwo\ (\RFour) jets.
 
\subsection{Observables}
 
The \bjet\ cross-section in \pp collisions  is defined as
\begin{equation}
\frac{\textrm{d}^{2}\sigma_{\bjet}}{\textrm{d}\pt \textrm{d}y} = \frac{1}{L_\textrm{int}\mathcal{B}}\frac{N^\textrm{Unfol.}_\textrm{\bjet}}{\Delta\pt\Delta y }\ ,
\label{eq:Xsecpp}
\end{equation}
where $L_\textrm{int}$ is the integrated luminosity and $\mathcal{B}$ is the branching ratio which includes
direct $(b\rightarrow\mu)$ and
cascade $(b\rightarrow c\rightarrow\mu)$ semileptonic decays.  The
value of $\mathcal{B}$ is
$(20.6\pm0.6)$\%~\cite{ParticleDataGroup:2020ssz}. The quantity $N^\textrm{Unfol.}_\textrm{\bjet}$ is the
\bjet\ yield after the unfolding and acceptance correction described above,
and $\Delta\pt$ and $\Delta y$ are the widths of the \pt and $y$ bins.
Following the same approach, the per-event yield of \bjets\ measured in \pbpb\
collisions is calculated as
\begin{equation}
\left. \frac{1}{N_\textrm{evt}}\frac{\textrm{d}^{2}N_\textrm{\bjet}}{\textrm{d}\pt \textrm{d}y}\right|_\textrm{cent} = \left.\frac{1}{N_\textrm{evt}\mathcal{B}}\frac{N^\textrm{Unfol.}_\textrm{\bjet}}{\Delta\pt\Delta y}\right|_\textrm{cent},
\label{eq:PeYHI}
\end{equation}
where $N_\textrm{evt}$ is the number of MB events in the centrality class `cent'.
The inclusive jet cross-section and per-event yields are
defined in the same way as in Eq.~(\ref{eq:Xsecpp}) and Eq.~(\ref{eq:PeYHI}) but without the $\mathcal{B}$ factor.

% End of text imported from the .//chapters/analysis.tex input file

\section{Systematic uncertainties}
\label{sec:systematics}

% The next lines are included from the .//chapters/systematics.tex input file
The systematic uncertainties of the jet cross-sections and per-event yields
common to both the inclusive jet and the \bjet\ measurements arise from the jet
energy scale and resolution,
the unfolding procedure, luminosity (\pp\ only), and
$\langle \TAA \rangle$ determination (\pbpb\ only).
There are additional contributions to the \bjet\ measurements from the modelling of the \ptrel\ distributions,
the fitting procedure, the branching ratio of \bhadrons\ to
muons, and the muon performance.
The uncertainties are discussed in detail below.
 
The systematic uncertainty of the jet energy scale (JES) has five parts. First,
there is a centrality-independent baseline component that is determined from in situ
studies of the calorimeter response for jets reconstructed with the procedure used in 13~\TeV\ \pp\ collisions~\cite{PERF-2011-03,Aaboud:2017jcu},
including an additional term for \bjet\ energy scale~\cite{PERF-2011-03}.
The second is a centrality-independent component which
accounts for the relative energy scale difference between the jet reconstruction procedure
used in this paper and that in 13~\TeV\ \pp\ collisions~\cite{ATLAS:2015twa}. Potential inaccuracies in the MC  description
of the relative abundances of jets initiated by quarks and gluons and of the calorimetric response to quark and gluon jets
are accounted for by a third, JES flavour,
component; this component is estimated independently
for the inclusive-jet and \bjet energy scale by varying the quark and gluon fractions from
their \pythiaeight\ values to those extracted from \herwig.
The fourth, centrality-dependent, component (in \pbpb\ collisions only) accounts for a different structure~\cite{ATLAS:2018bvp}, and
possibly a different detector response, for jets in \PbPb\ collisions that is not modelled by the simulations.
It is evaluated by the method used
for 2015 and 2011 data~\cite{ATLAS:2015twa} that compares the calorimetric
jet \pt\ and the sum of the transverse momentum of charged particles
within the jet in data and MC samples. The size of the centrality-dependent uncertainty in the JES reaches 1.2\% in the most central collisions; it is smaller
for more peripheral collisions.
The systematic uncertainties from
the JES discussed above are derived for $R=0.4$ jets and applied to both \RFour\ and \RTwo\
jets.
An additional component applies only to \RTwo\ jets and
accounts for a potential uncertainty difference between  $R = 0.4$ and $R = 0.2$ jets.
The uncertainty is assessed by comparing the ratio of \RTwo\ jet \pt\ to \RFour\ jet \pt\ in data and simulations.
 
The uncertainty due to the jet energy resolution (JER) is evaluated by repeating the unfolding procedure with modified response matrices, where an additional contribution is added to the resolution of the reconstructed \pT\ using a Gaussian smearing procedure.
The smearing factor is evaluated using an in situ technique in 13~\TeV\ \pp\ data that involves studies of dijet
\pT\ balance~\cite{Aad:2012ag,Aaboud:2019ibw}.
Additionally, an uncertainty is included to account for differences between the tower-based jet reconstruction and the jet reconstruction used in
analyses of 13~\TeV\ \pp\ data, as well as differences in calibration procedures.
Similarly to the JES, an additional uncertainty in the JER accounting for differences between $R = 0.2$  and $R = 0.4$ jets is added.
The resulting uncertainty from the JER is symmetrized.
 
Uncertainties related to muons are associated with the trigger and reconstruction efficiency measurement.
The systematic uncertainties are estimated by varying the tag-and-probe method
as described in Refs.~\cite{ATLAS:2016lqx,Aad:2020uyd}.
Additionally, the statistical uncertainty of the factor used to correct the centrality dependence of
the trigger efficiency
in Pb+Pb collisions and the  difference between the data-driven reconstruction efficiencies
in \pbpb\ and \pp\ collisions
are taken as muon systematic uncertainties that apply only to \pbpb\ collisions.
 
The uncertainties in modelling the \ptrel\ distributions come from several sources, which are common to \pp and \pbpb\ collisions.
For the fraction of \bhadrons\ which
arise from gluon splitting, the analysis uses that from \pythiaeight\ simulation as the central value. In \pp collisions,
the gluon-splitting fraction
is reweighted to the value obtained from the \herwig\ sample~\cite{ATLAS:2021agf}.
In \pbpb\ collisions,
there is no information about the modification of the gluon-splitting contribution to
\bjets.  Additionally, the fragmentation functions of \bjets\ in \pbpb\ collisions have not
been measured.
In order to cover both of these effects, the gluon-splitting fraction is conservatively varied between zero and 100\%.
For the \cjets, the gluon-splitting uncertainty is estimated by varying the
\pythiaeight\ value of the gluon-splitting fraction
by a factor of two, based on measurements in Ref.~\cite{CMS:2019jis}.
The fraction of muons which arise from \bhadron\ decays which include
an intermediate \dmeson\ is varied in accord with Ref.~\cite{ParticleDataGroup:2020ssz}.
The fractions of the various \bhadron\ species are varied according to the world average values in Ref.~\cite{Amhis:2014hma}.
The fraction of c-baryons was varied according to ALICE measurement~\cite{ALICE:2021bib} with negligible impact on the results.
The modelling of the muon momentum in the \bhadron\ rest frame is crucial for the \ptrel\ method.
The modelling in \pythiaeight\ is used for the central values.
The \pythiaeight\ distributions were compared
with the measurement from DELPHI~\cite{Abdallah:2005cx}
and the difference between \pythiaeight\ and the DELPHI measurement is used as a
systematic uncertainty.
The uncertainty in the \ljet\
template is evaluated by using muons with a distance of closest
approach to the collision vertex smaller than 0.01~mm,
to minimize the contribution from heavy-flavour jets, and remaking the templates.
In \RTwo\ jets in \pp  collisions, the dominant components of modelling uncertainties
stem from the \ljet\ template and the fraction of \bhadrons;
for \RFour\ jets, the \ljet\ template uncertainty dominates.
For \pbpb\ collisions, the dominant modelling uncertainties
stem from the \ljet\ template and gluon-splitting components.
 
The uncertainty in the unfolding procedure is determined in all cases by
constructing response matrices
from the MC distributions without the reweighting factors that are used to match
the MC distributions to those in data. The uncertainty due to using a particular
MC model in unfolding \bjets\ is addressed by reweighting
the 2D (\ptjet$,$\ptmu) distribution in \pythiaeight\ to that observed in \herwig.
The inclusive jet analysis has an uncertainty associated with
the non-closure of the unfolding procedure when the MC sample is divided
and one portion is used in place of the data and the other portion
used to generate a response matrix;
in the \bjet\ analysis the closure was found to be consistent with
unity within the statistical uncertainties and no uncertainty is added.
 
Uncertainties from the finite size of the MC samples
are combined with the data statistical uncertainties.
A systematic uncertainty associated with the template
fit procedure is addressed by allowing the templates to
deform in \pbpb\ collisions by
convolving the nominal template with a Gaussian function
where the width parameter is free.
Deformations of the templates account for mismodelling due to the large UE,
such as via the jet position resolution, muon momentum resolution, or other effects.
A similar uncertainty was applied in Ref.~\cite{ATLAS:2021xtw}.
The small non-closure of the fitting procedure, when the \fb\
extraction is tested in simulations,
is also included in the
systematic uncertainties.
 
\begin{figure}[htbp]
\centering
\includegraphics[width=0.45\textwidth]{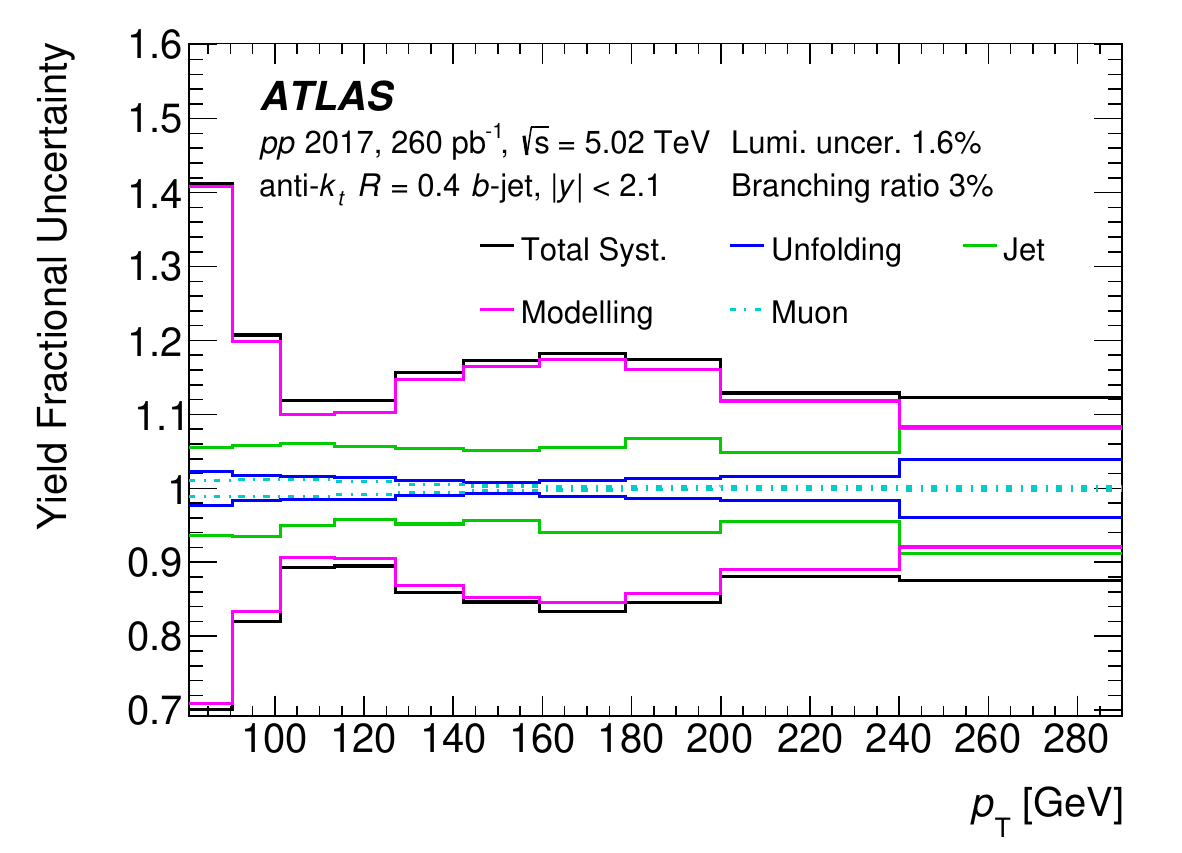}
\includegraphics[width=0.45\textwidth]{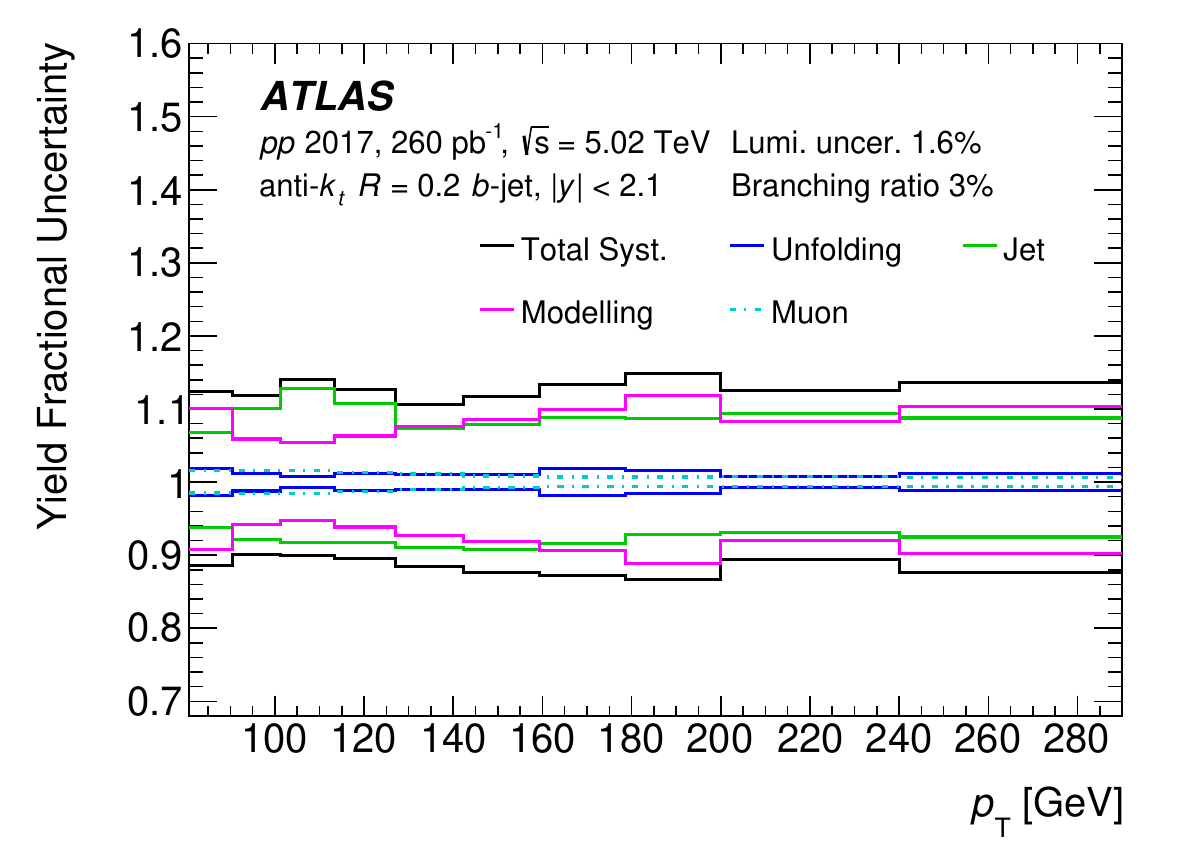}\\
\includegraphics[width=0.45\textwidth]{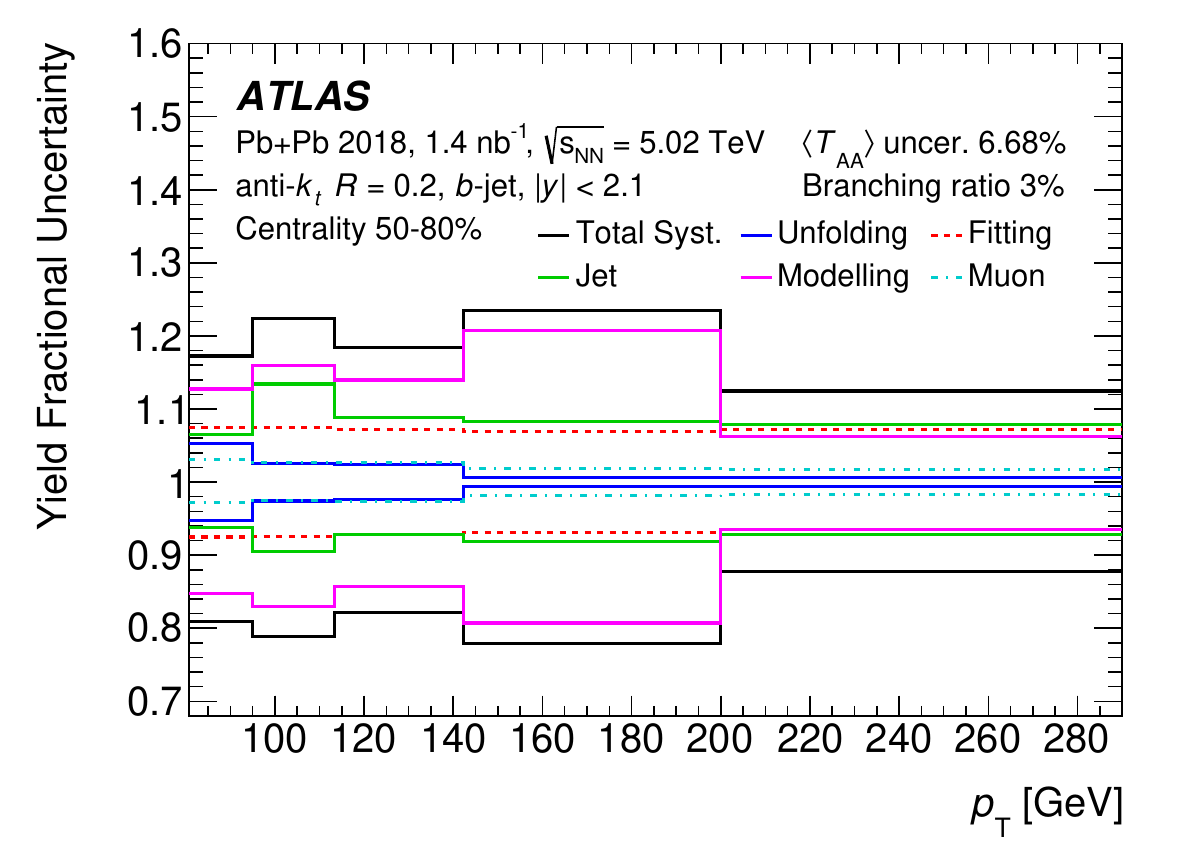}
\includegraphics[width=0.45\textwidth]{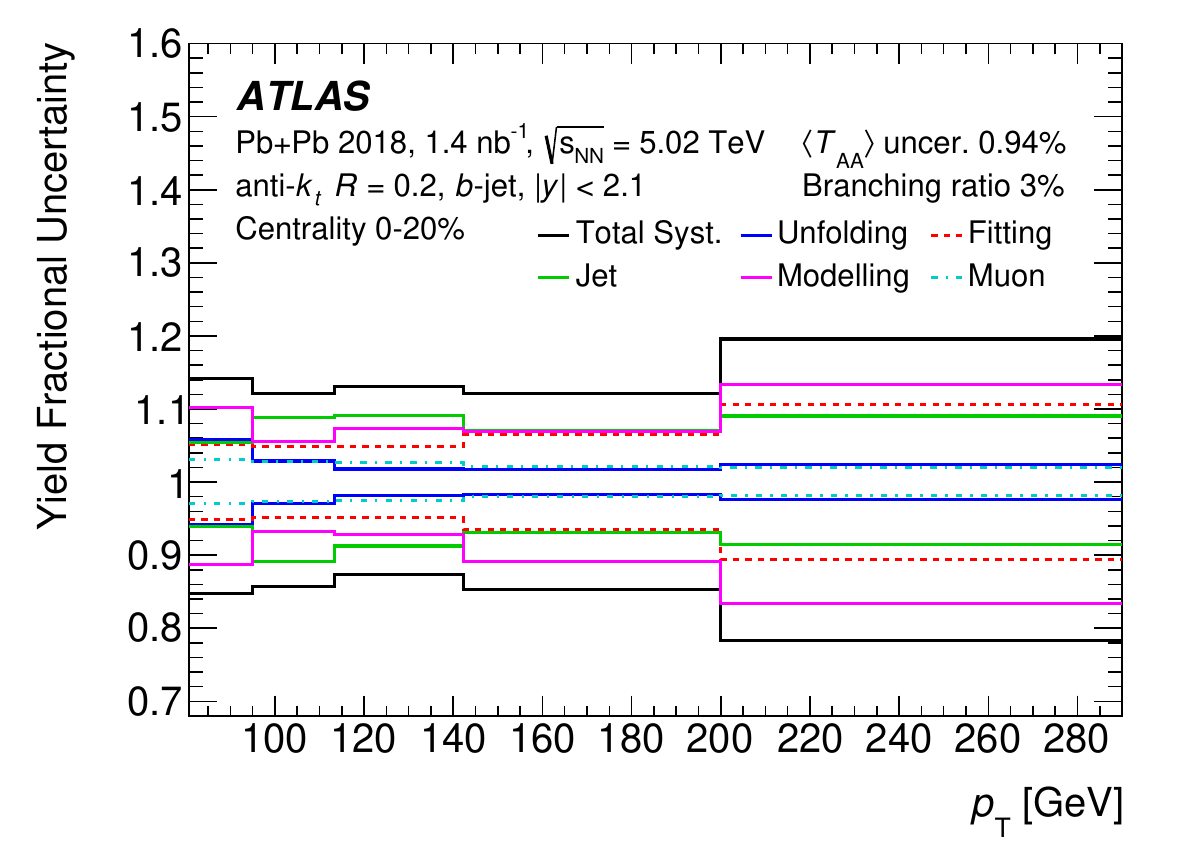}\\
\caption{The relative systematic uncertainties for several categories,
as a function of $b$-jet \pt\ for
(top) the \pp cross-section for \RFour\ (left) and \RTwo\ (right) jets, and
(bottom) the Pb+Pb per-event-yield in peripheral (left) and central (right) collisions
for \RTwo\ jets.}
\label{fig:sysXsec}
\end{figure}

For each uncertainty component discussed above, the entire analysis procedure
is repeated with the variation under consideration and the resulting changes
are added in quadrature to form the total systematic uncertainty of the measurement.
A summary of the systematic uncertainties for the
\RFour\ and \RTwo\ \bjet cross-sections is shown in Figure~\ref{fig:sysXsec}.
For the \RFour\ jets, the largest contribution is the modelling component, while for
\RTwo\ jets, the jet and modelling components have
similar magnitudes.
 
The integrated luminosity determined for 2017 \pp data was calibrated using data from dedicated
beam-separation scans, also known as van der Meer scans~\cite{ATLAS:2016fhk}.
Sources of systematic uncertainty similar to those examined in the 2012 \pp
luminosity calibration~\cite{ATLAS:2016fhk}
were studied in order to assess the systematic uncertainties for the 2017 data.
The combination of these systematic uncertainties results in a relative uncertainty of 1.6\%.
The uncertainty of the mean nuclear thickness function arises from geometric modelling uncertainties (nucleon--nucleon inelastic cross-section,
Woods--Saxon parameterization of the nucleon distribution) and the uncertainty in the fraction of selected inelastic Pb+Pb collisions.
The values of these uncertainties are taken from Ref.~\cite{Aaboud:2018twu}.
The  branching ratio for \bhadrons\ into muons is $(20.6 \pm 0.6)\%$ and is
taken from Ref.~\cite{ParticleDataGroup:2020ssz}.

\begin{figure}[htbp]
\centering
\includegraphics[width=0.45\textwidth]{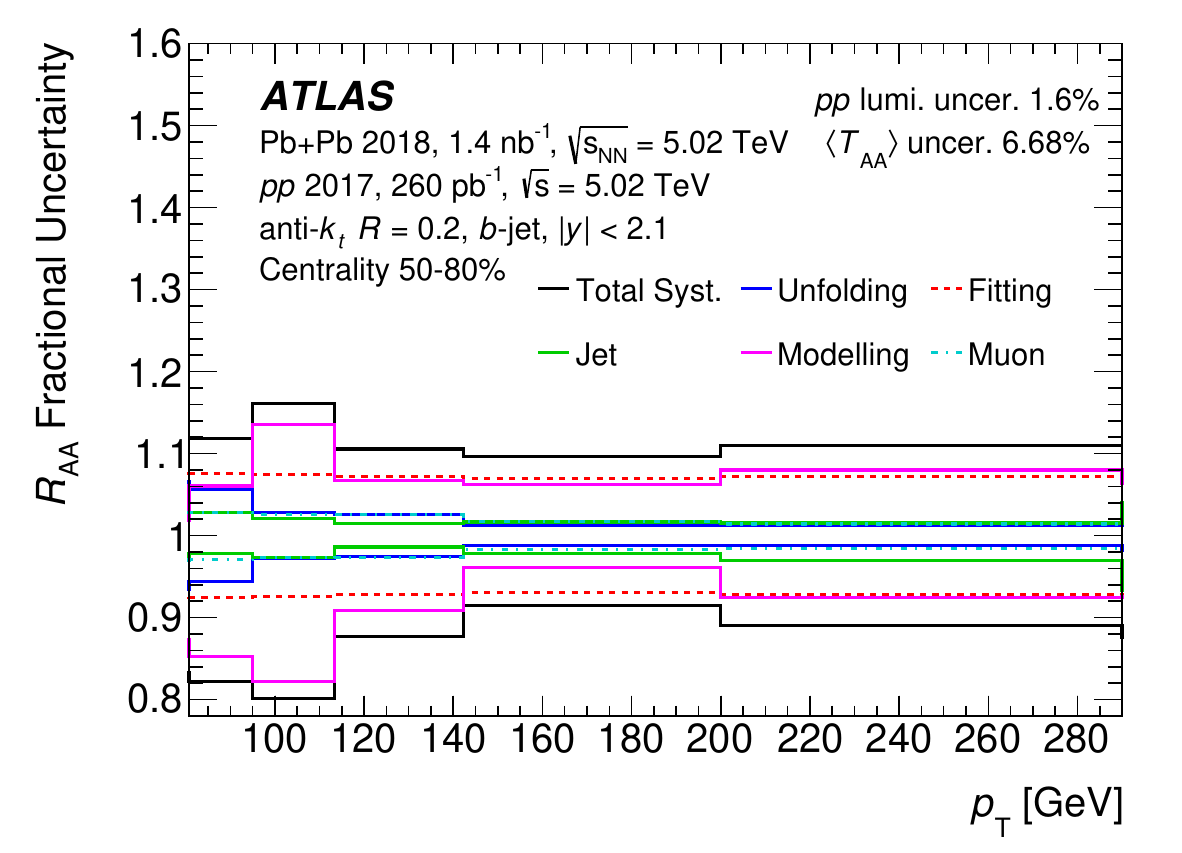}
\includegraphics[width=0.45\textwidth]{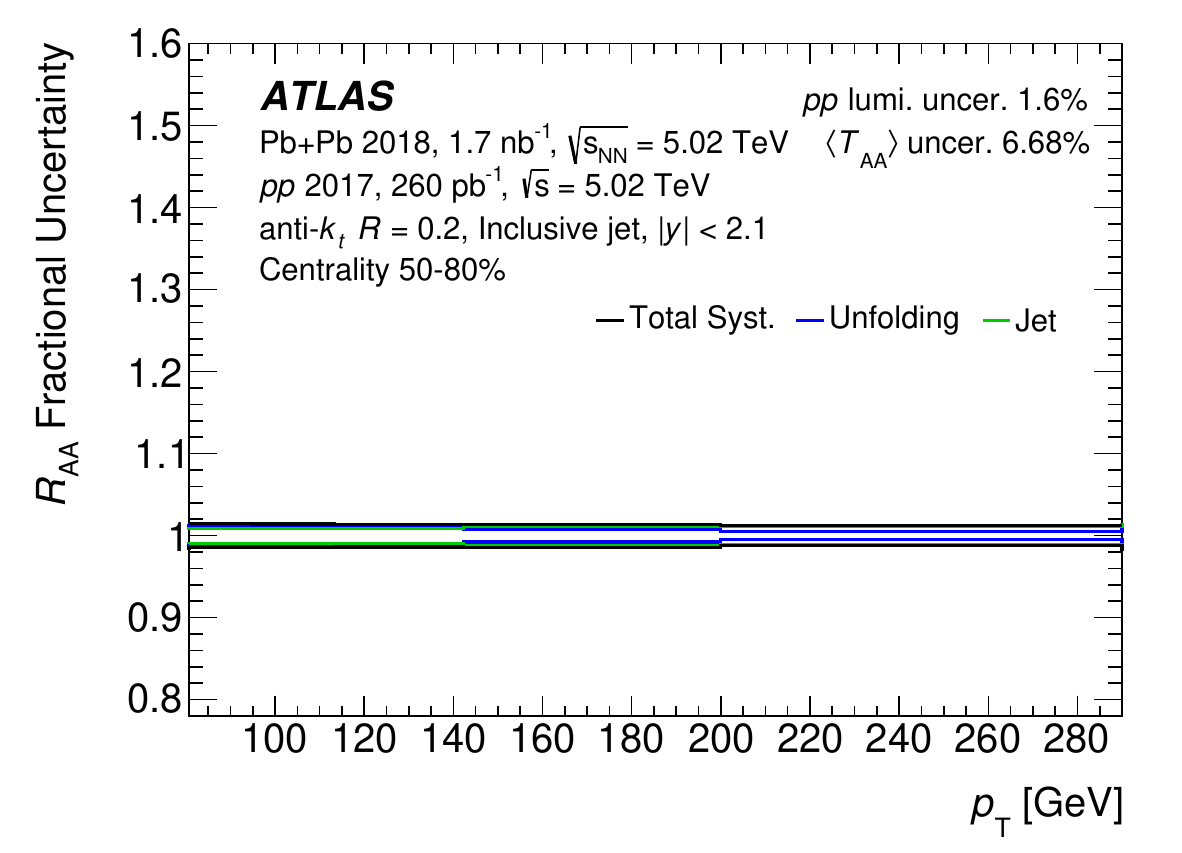}\\
\includegraphics[width=0.45\textwidth]{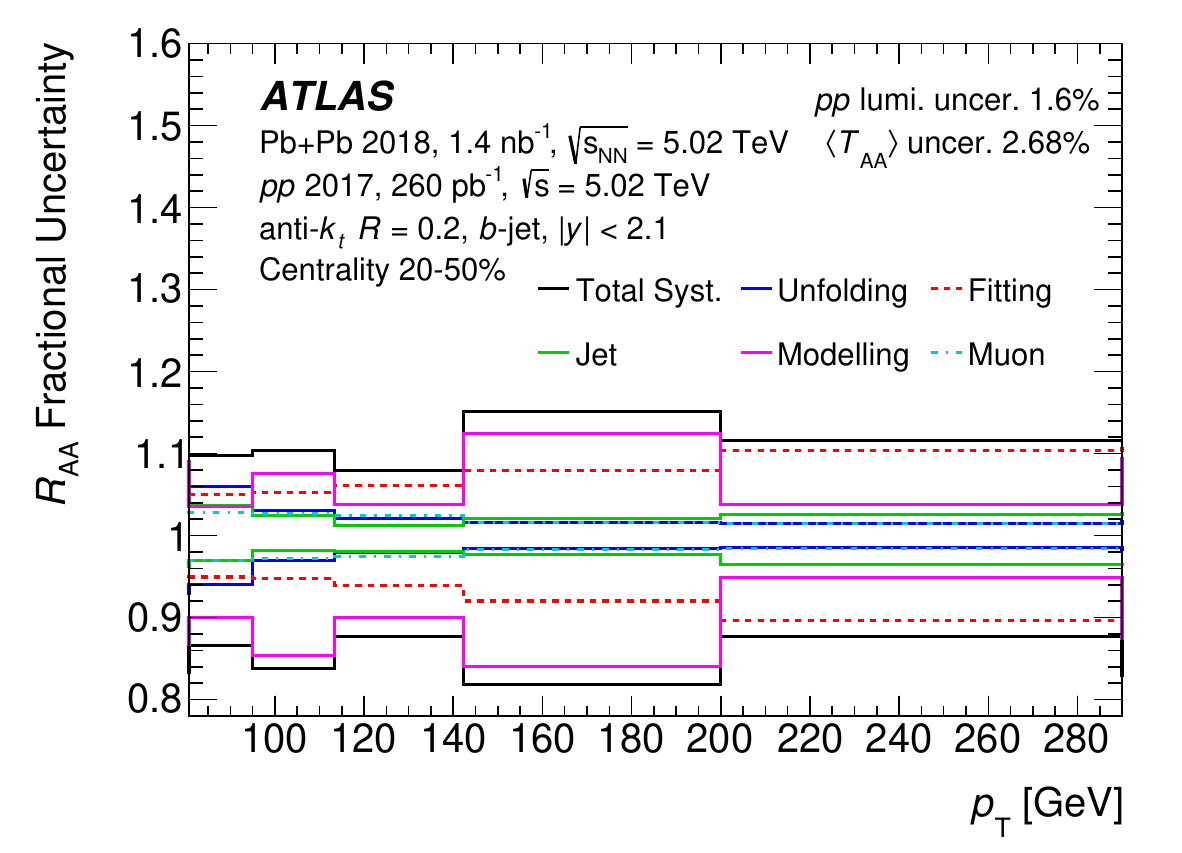}
\includegraphics[width=0.45\textwidth]{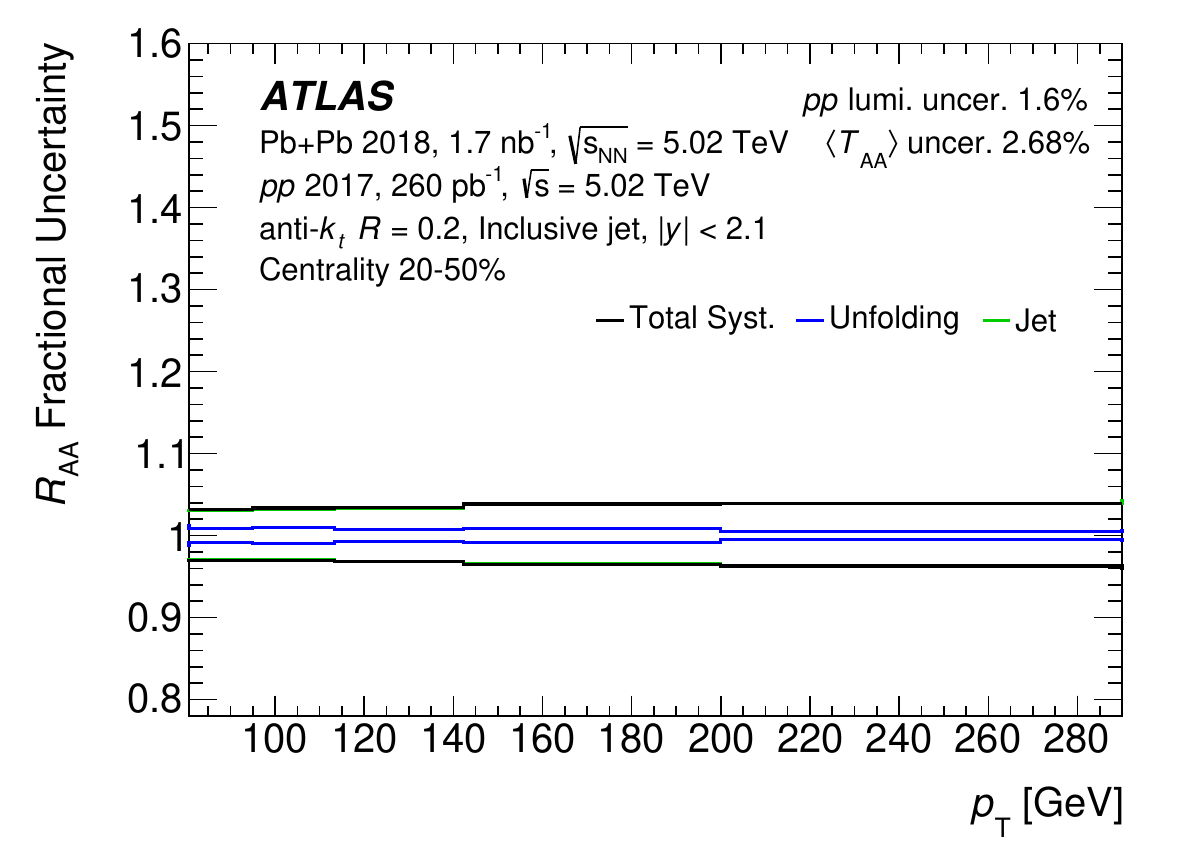}\\
\includegraphics[width=0.45\textwidth]{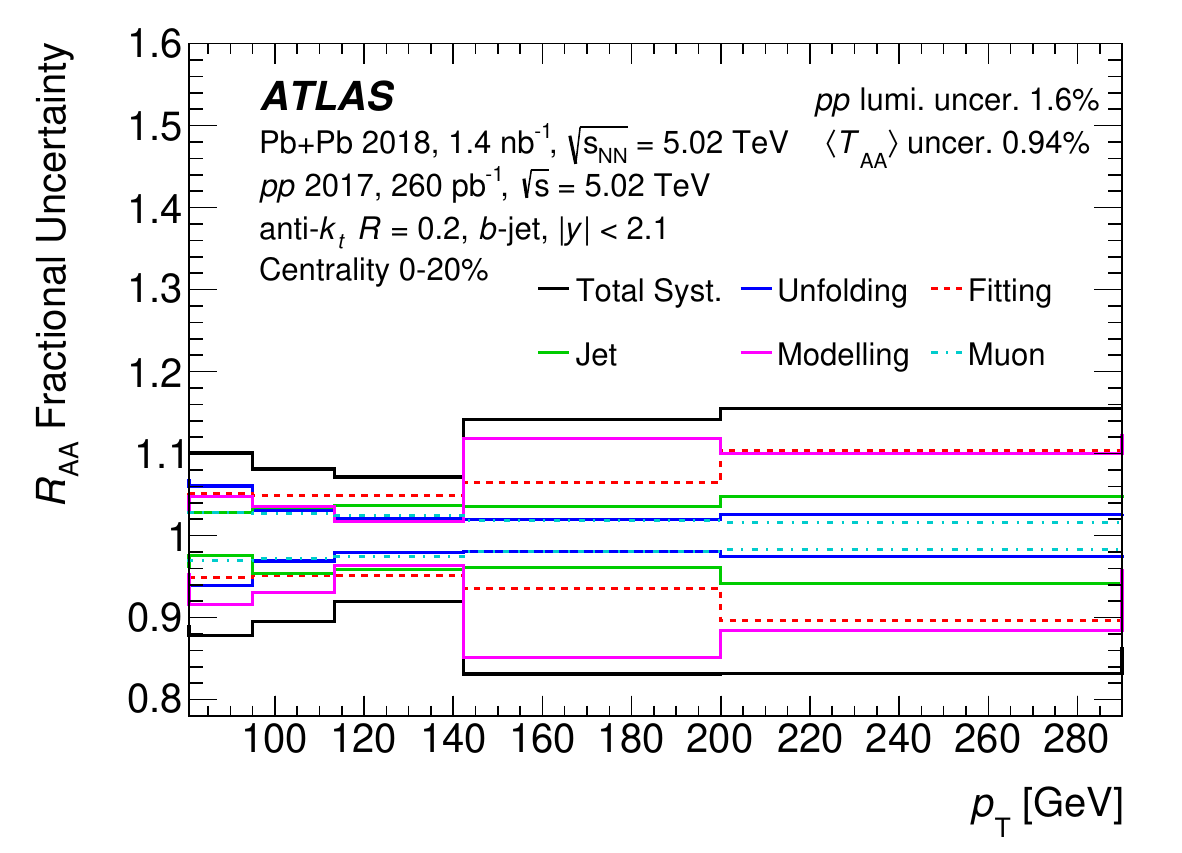}
\includegraphics[width=0.45\textwidth]{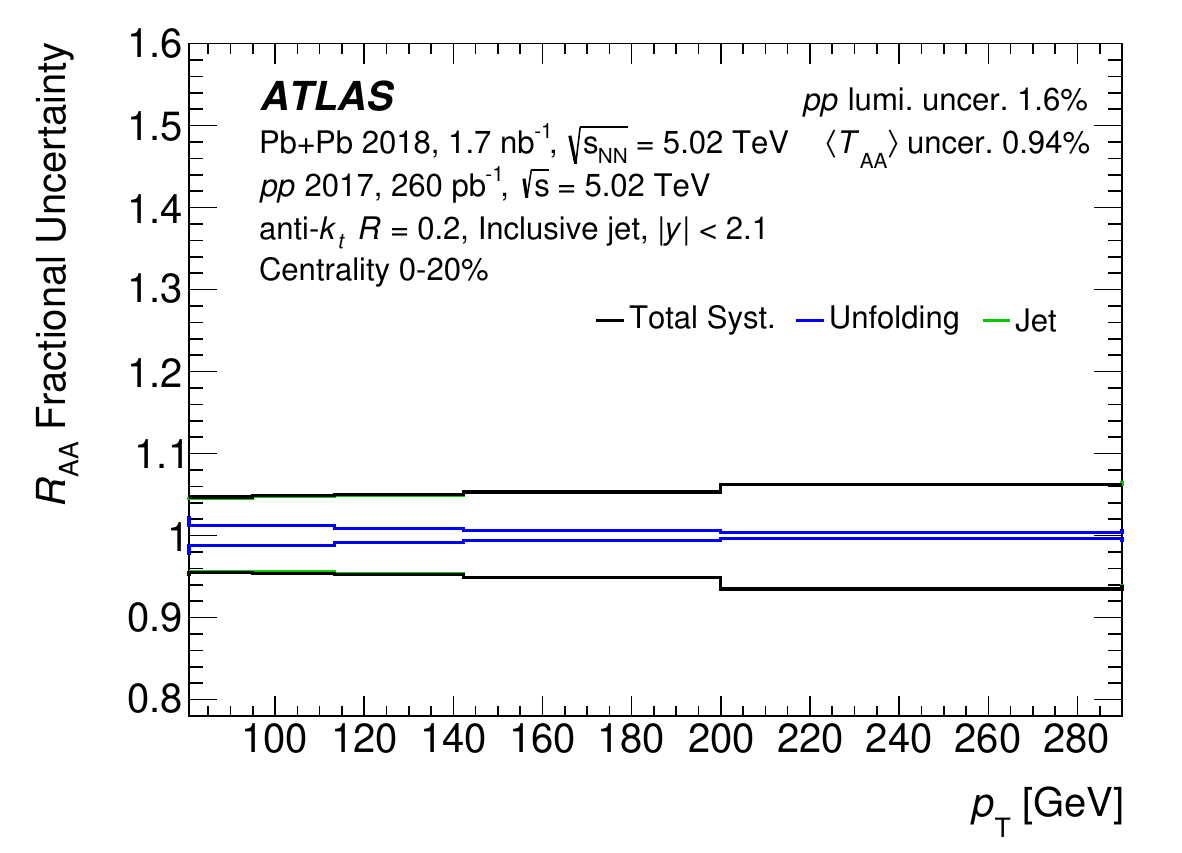}
\caption{Relative size of systematic uncertainties for several categories,
as a function of jet \pt\ for
\bjet \RAA\ (left column) and  inclusive jet \RAA\ (right column),
in  peripheral (top row),
semi-central (middle row) and  central (bottom row) \pbpb\ collisions.}
\label{fig:sysRAA}
\end{figure}

\begin{figure}[htbp]
\centering
\includegraphics[width=0.55\textwidth]{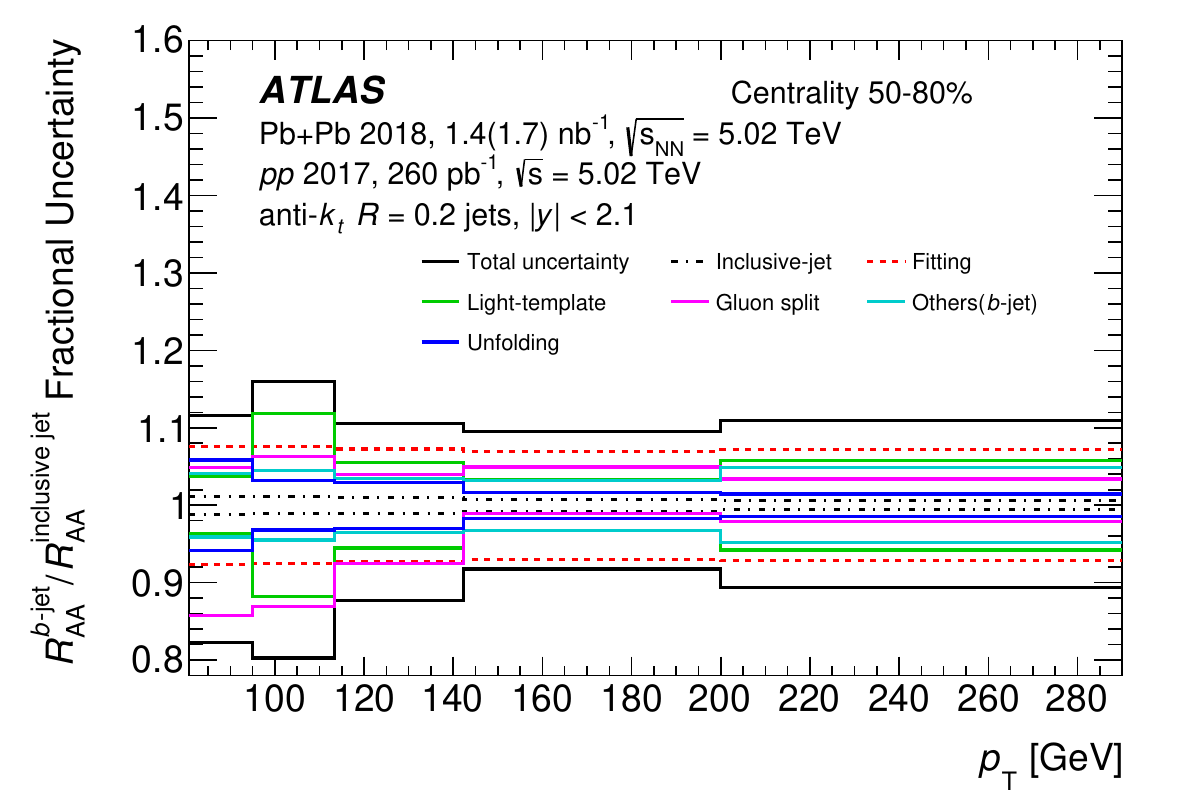}\\
\includegraphics[width=0.55\textwidth]{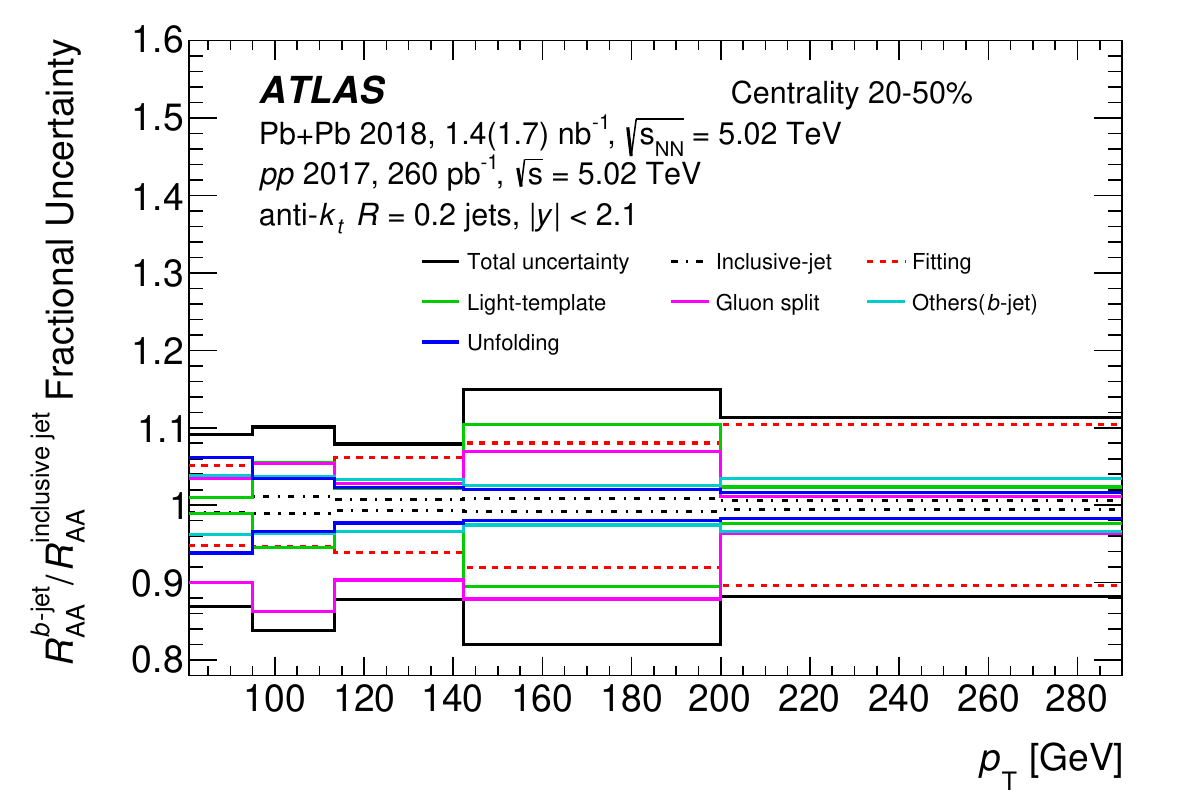}\\
\includegraphics[width=0.55\textwidth]{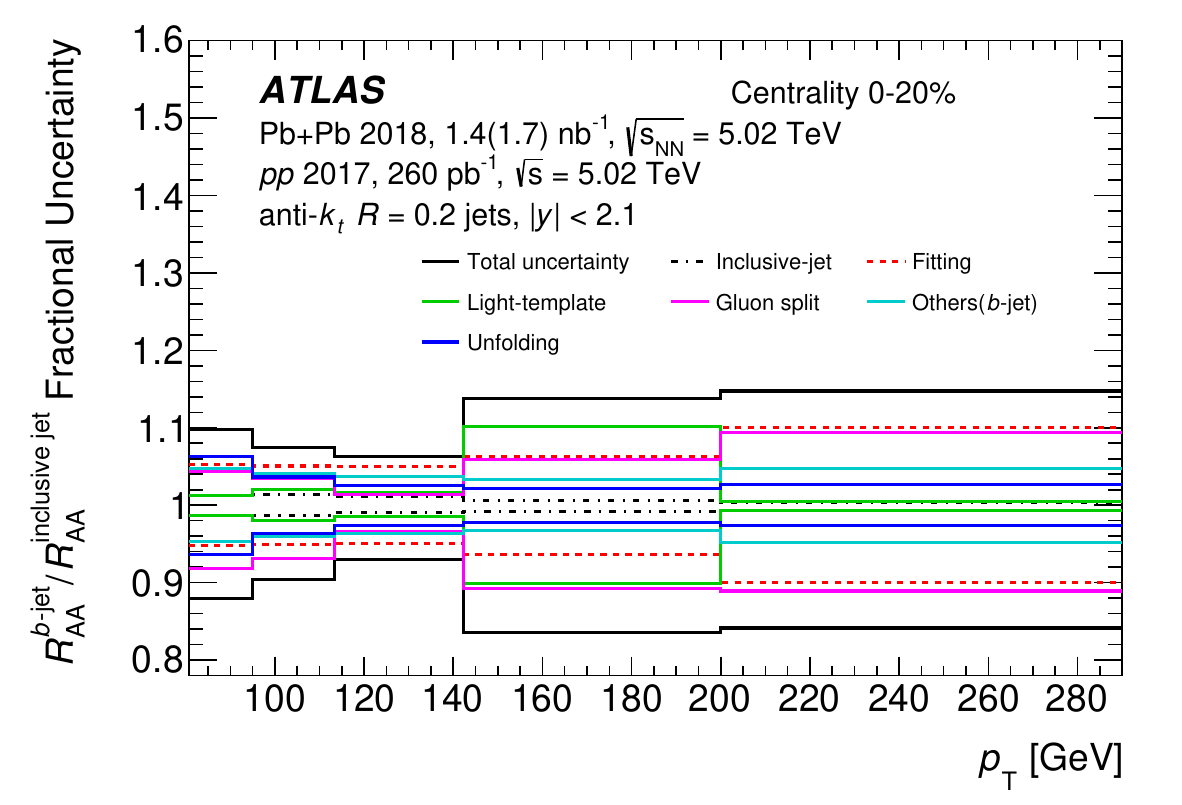}
\caption{Relative size of systematic uncertainties of the ratio
$\RAA^{\bjet}/\RAA^\textrm{inclusive jet}$
shown for the most relevant components,
as a function of jet \pt\ for
50--80\% (top),
20--50\% (middle), and 0--20\% (bottom) centrality \pbpb\ collisions.}
\label{fig:sysRAA_ratio}
\end{figure}

The uncertainties which are common to \pp\ and \pbpb\ collisions are treated as correlated when
determining the uncertainty in the \RAA\ value, with the
exception of gluon splitting, where the uncertainties
account for possible production and fragmentation mechanism differences
between \pp and \pbpb\
collisions, and the unfolding.
Similarly, the uncertainties which are common to inclusive jets
and \bjets\ cancel out when ratios of cross-sections or \RAA\ values are taken;
the remaining uncertainties are the JES flavour component,
the \bjet-specific JES uncertainty, and the unfolding.
A summary of the systematic uncertainties for inclusive jet and \bjet \RAA\
is shown in Figure~\ref{fig:sysRAA}.
The uncertainties in the $\RAA^{\bjet}/\RAA^\textrm{inclusive jet}$ ratio
are shown in Figure~\ref{fig:sysRAA_ratio}; the uncertainties in the
gluon-splitting contribution and deformation of the templates in \pbpb\ collisions dominate in
most cases.
 
% End of text imported from the .//chapters/systematics.tex input file

\FloatBarrier
\section{Results}
\label{sec:result}

% The next lines are included from the .//chapters/results.tex input file
\subsection{Cross-section in \pp\ collisions}
 
\begin{figure}[htbp]
\centering
\includegraphics[width=0.49\textwidth]{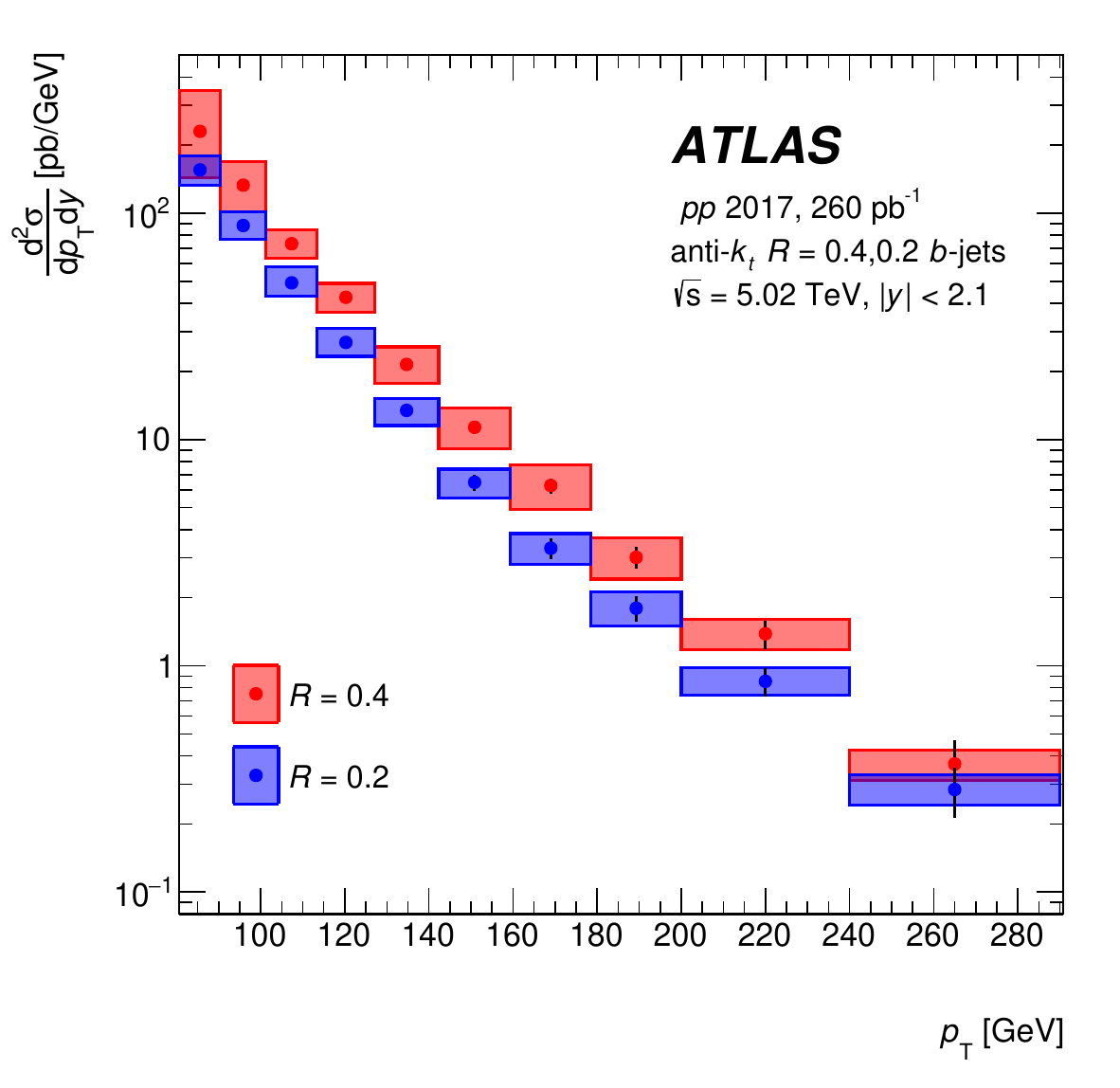}
\includegraphics[width=0.49\textwidth]{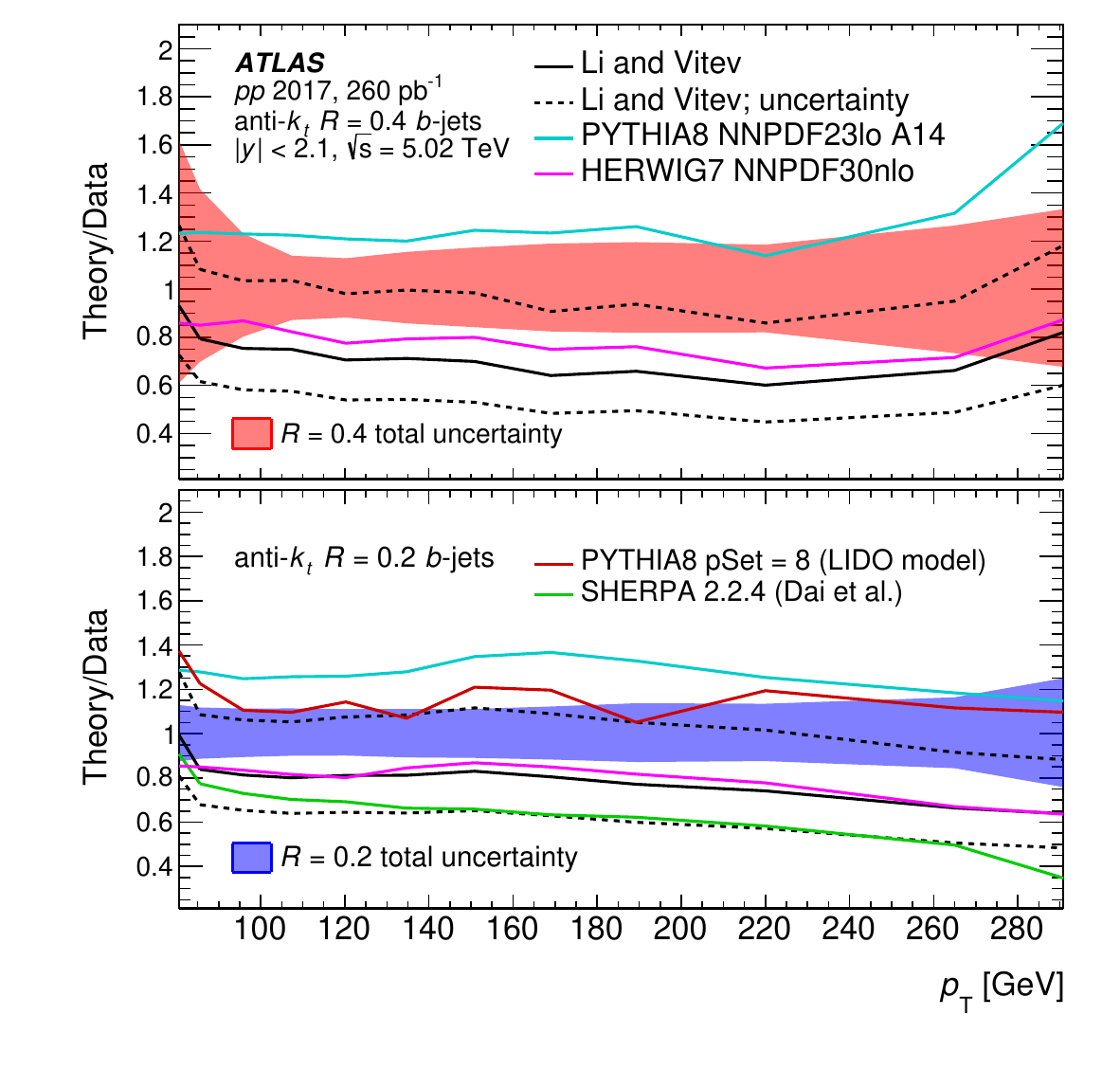}
\caption{(left) Differential cross-section for \bjet\ production
for \RTwo\ and \RFour\ jets with $|y|<2.1$
as a function of \pt\
in 5.02~\TeV\ \pp data.
(right) Ratio of the predictions to the measured \textit{b}-jet cross-section in \pp collisions at 5.02~\TeV\
for \RFour\ (top) and \RTwo\ (bottom) jets.
The \RFour\ jets are compared with calculations from Ref.~\cite{Li:2018xuv} and \RTwo\ jets are compared with calculations from Refs.~\cite{Li:2018xuv,Ke:2018jem,Ke:2018tsh,Dai:2018mhw}.
Both cross-sections are compared with \pythiaeight\ and \herwig\ calculations.
The bands around unity represent the total uncertainty of the data.
The text provides additional discussion.}
\label{fig:Xsec5pp_theory}
\end{figure}

Figure~\ref{fig:Xsec5pp_theory}  shows the \textit{b}-jet cross-section
as a function of \textit{b}-jet \pt in \pp collisions at 5.02 \TeV\ for \RTwo\ and \RFour\ jets with $|y|<2.1$.
Additionally, \bjet\ cross-sections for both \RTwo\ and \RFour\ jets are
compared with theory and generator calculations.
The data is in good agreement with the calculations by Li and Vitev~\cite{Li:2018xuv}
for both \RFour\ and \RTwo\ jets. This calculation is based on
semi-inclusive jet functions where the cross-section is expressed in terms of
the parton distribution functions, the hard kernel and the jet functions.
The jet functions have terms in $\ln(R)$, which are resummed.
The $\SHERPA\,2.2.4$ calculations~\cite{Dai:2018mhw} for \RTwo\ jets underestimate
the measured \bjet\ cross-sections by an amount which increases with \pt.
A calculation~\cite{Ke:2018jem,Ke:2018tsh} which is based
on \pythiaeight\ where the PDFs and $\alpha_{s}(\textrm{M}_{Z})$ is set with $\mathrm{pSet} = 8$~\cite{Sjostrand:2014zea}
is approximately consistent with
the upper edge of the uncertainty band of the \RTwo\ jet yields.
The \pythiaeight\ calculation using the \textsc{NNPDF23lo} PDF~\cite{Ball:2012cx} and the A14 tune~\cite{TheATLAScollaboration:2014rfk}
is 20--30\% higher than data for both \RFour\ and \RTwo\ jets.
The \herwig\ calculations~\cite{Bellm:2015jjp}
using the \textsc{NNPDF30nlo} PDF  underestimate
the measured cross-sections at both jet radii.
There is no uncertainty estimate for the generator calculations, and the sensitivity of the
results to the choice of parameter values has not been investigated.
 
\begin{figure}[htbp]
\centering
\includegraphics[width=0.49\textwidth]{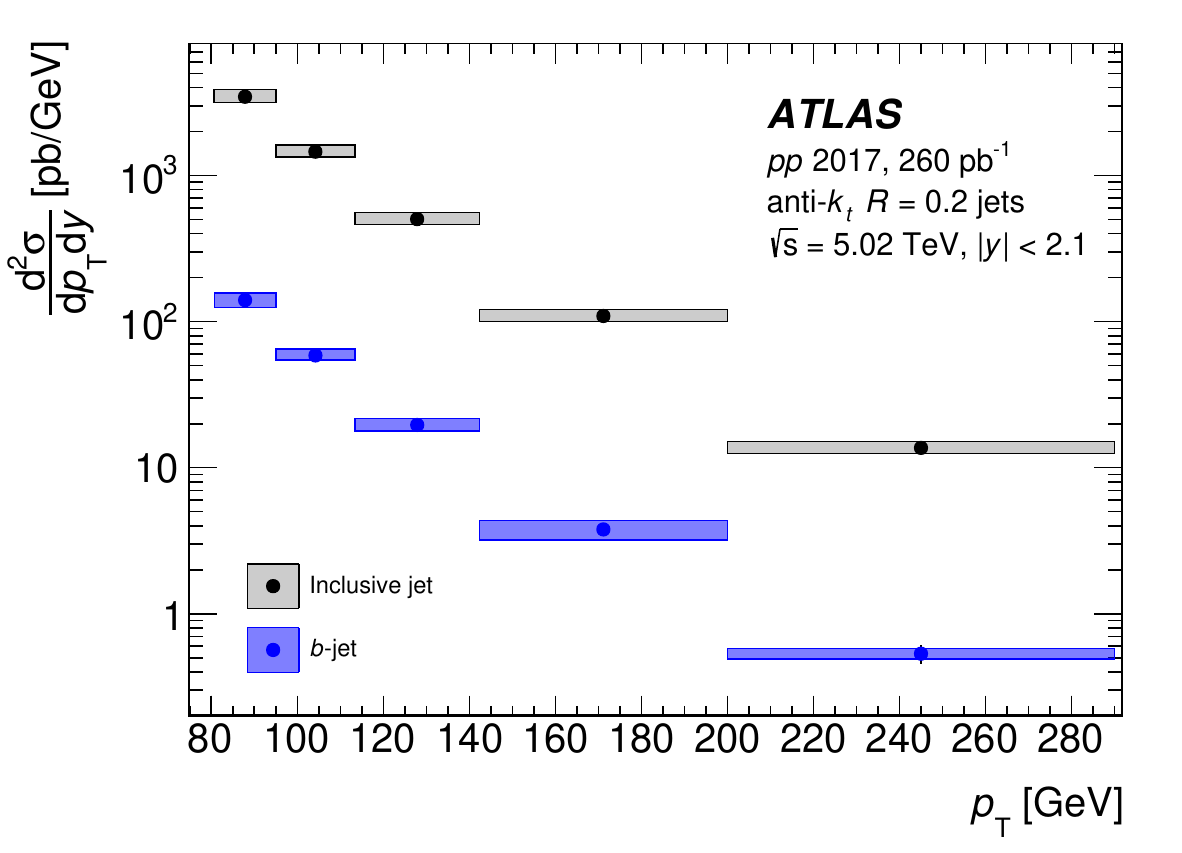}
\includegraphics[width=0.49\textwidth]{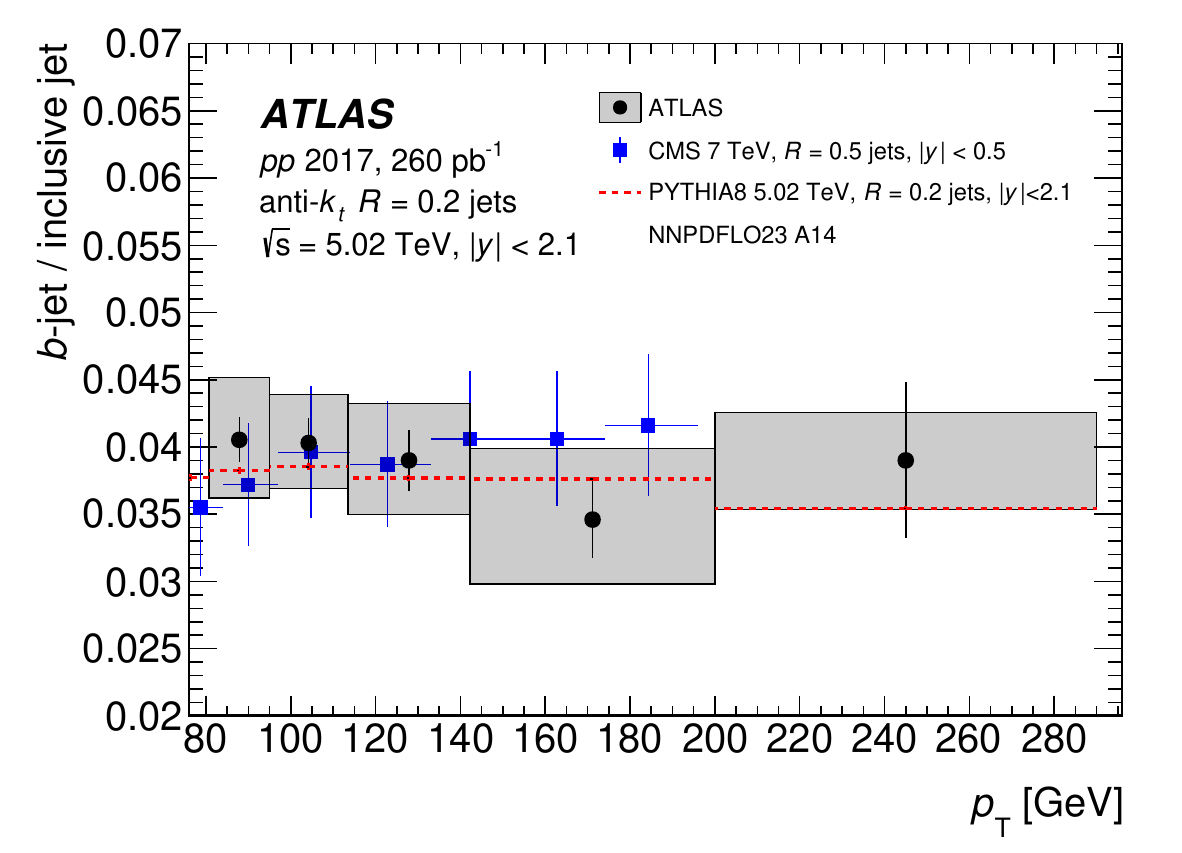}
\caption{(left) Cross-section of \RTwo\ \bjet and inclusive jet production in \pp collisions at 5.02~\TeV,
and (right) the \bjet to inclusive jet cross-section ratio, together
with \pythiaeight\ simulation and measurements
from the CMS Collaboration at 7~\TeV\ for \RFive\ jets~\cite{Chatrchyan:2012dk}.
For the current measurement,
the boxes represent the systematic uncertainties and the error bars represent the statistical uncertainties.
For the CMS data the bars represent the total uncertainty.}
\label{fig:Xsecpp_inclusive}
\end{figure}

Figure~\ref{fig:Xsecpp_inclusive} shows
a comparison of the \bjet\ and inclusive jet cross-sections for \RTwo\ jets
and their ratio.
The figure also shows the same ratio from the \pythiaeight\ MC
sample; it agrees with the data.
The cross-section ratio measured at 7~\TeV\ for \RFive\ jets with $|y|<0.5$
by the CMS Collaboration~\cite{Chatrchyan:2012dk}
is also shown; the jet radii in this measurement are larger than in
the present analysis but the results are qualitatively similar.
No significant \pt\ dependence of the \bjet\ to inclusive jet cross-section
ratio is observed.
 
\subsection{Per-event yields and \RAA\ in \pbpb\ collisions}

\begin{figure}[!h]
\centering
\includegraphics[width=0.49\textwidth]{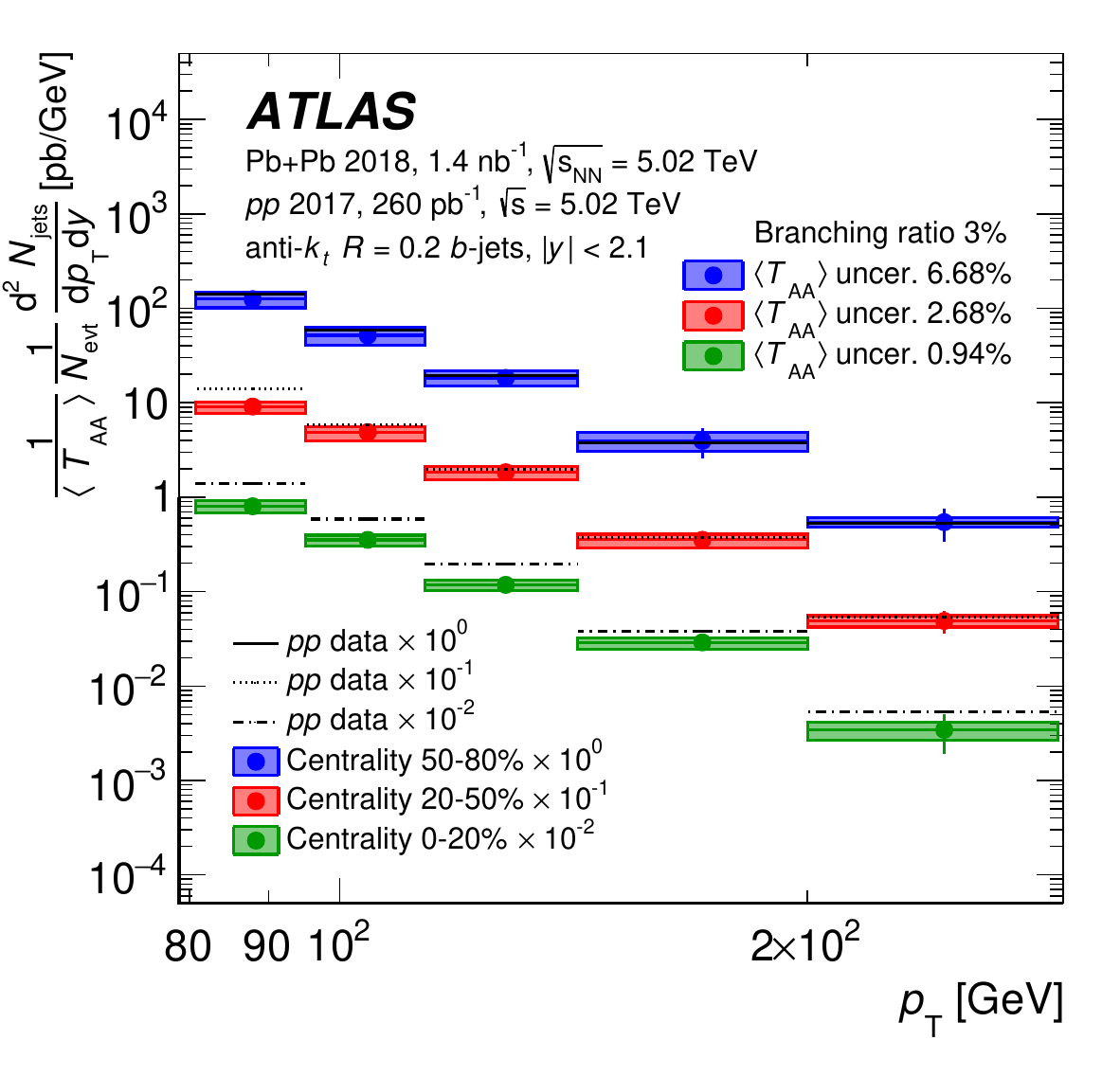}
\includegraphics[width=0.49\textwidth]{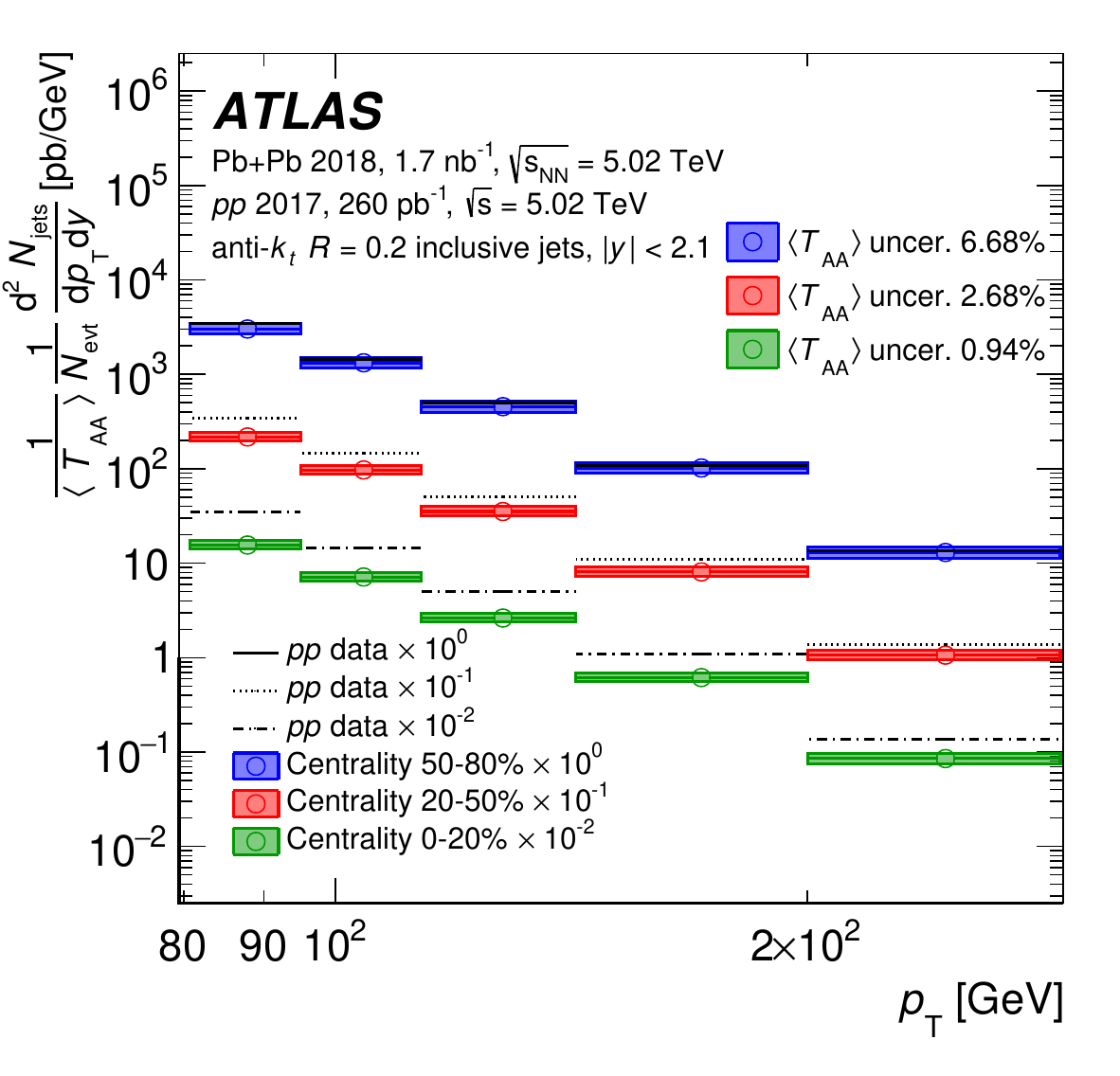}
\caption{
Per-event yields scaled by $\langle\TAA\rangle$ in \pbpb\ collisions
for three centrality classes for \RTwo\  \bjets (left) and  inclusive jets (right).
The boxes show the systematic uncertainties and the bars represent the statistical uncertainties.
The different centrality classes are offset by the factors shown on the plot for clarity.
The values of the \pp\ cross-sections are shown by the black lines; they are offset by the factors
shown on the plot.
The size of the \TAA\ uncertainties are noted in the legend.}
\label{fig:Xsec_PEY_syst}
\end{figure}

\begin{figure}[htbp]
\centering
\includegraphics[width=0.65\textwidth]{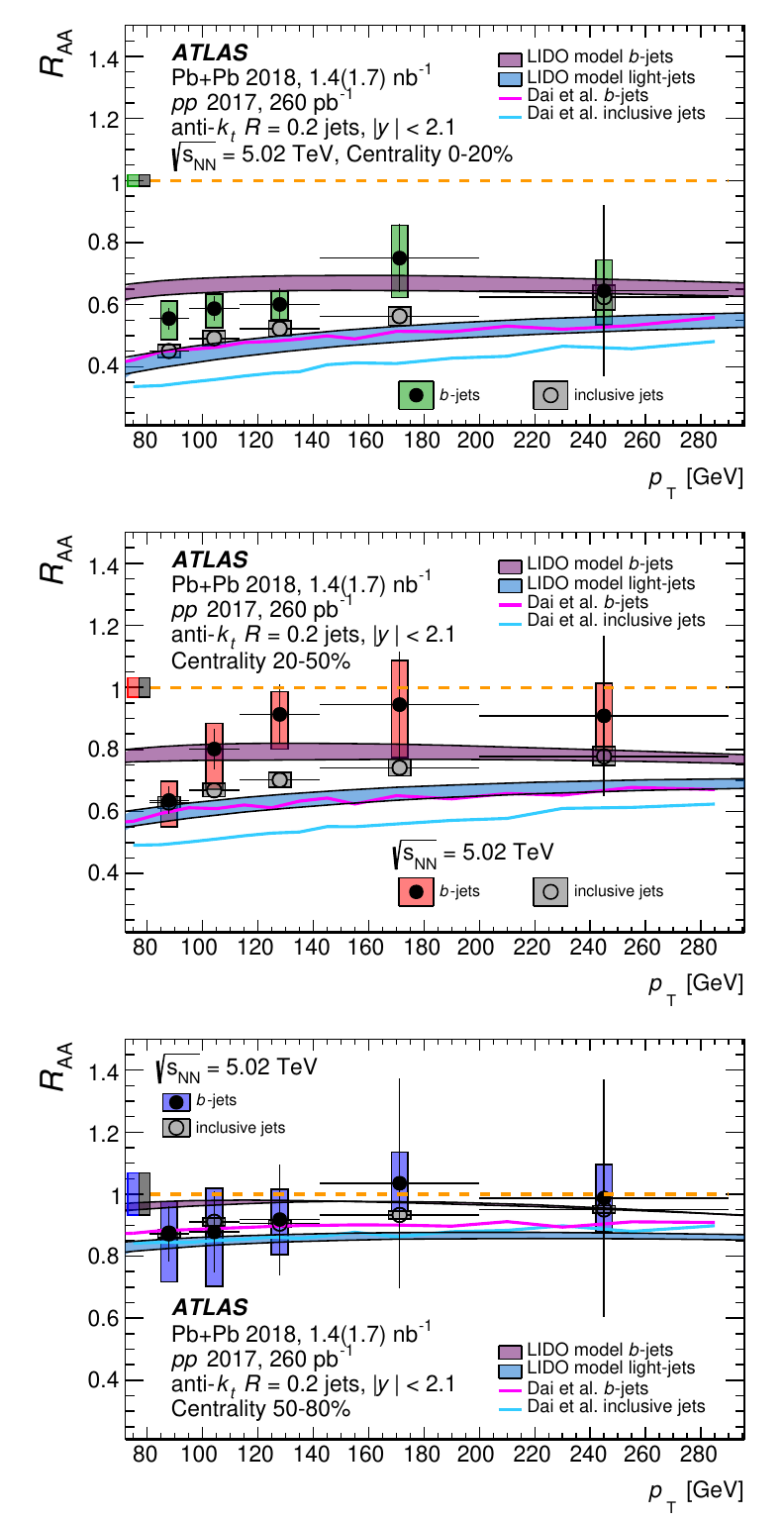}
\caption{\bjet\  (filled points) \RAA\
for three centrality classes compared with the inclusive jet
\RAA\ (open points) at 5.02~\TeV.
Both \RAA\ measurements
are compared with theory calculations~\cite{Ke:2018jem,Ke:2018tsh,Dai:2018mhw}.
The boxes represent the systematic uncertainties and the error bars
represent the statistical uncertainties.
For the LIDO calculation, the width of the band shows the variation of the $\mu_\textrm{min}$
parameter from $1.3\pi T$ (lower edge) to $1.8\pi T$ (upper edge).
The boxes at unity represent the scale uncertainties from $\langle\TAA\rangle$ and
the luminosity determination.}
\label{fig:RAA_light_b_comp}
\end{figure}

The inclusive jet and \bjet\ per-event yields in \pbpb\ collisions scaled by $\langle\TAA\rangle$
are shown in Figure~\ref{fig:Xsec_PEY_syst} for the three centrality classes used
in this analysis overlaid with the values of the \pp\ cross-sections.
Figure~\ref{fig:RAA_light_b_comp} shows a direct comparison of the inclusive jet and \bjet\ \RAA\
for each centrality class.
The \RAA\ values for both types of jets decrease going from peripheral to central collisions.
Both are consistent with unity in peripheral collisions.  In central and semi-central collisions
the \RAA\ central values for inclusive jets are lower than those for \bjets.
A difference between the slopes of the inclusive jet and \bjet\ differential cross-sections
could cause the \RAA\ values to differ between the two categories of jets;
however, the ratio of the \bjet\ to inclusive jet cross-section, shown in Figure~\ref{fig:Xsecpp_inclusive},
does not vary with \pt.
 
The \RAA\ values for both inclusive jets and \bjets\  are
compared with two theory calculations.
The first calculation is the LIDO model~\cite{Ke:2018jem,Ke:2018tsh,Ke:2020nsm},
which includes both energy loss and diffusion of heavy quarks.
The dead-cone effect for \bjets\
is included and the medium is implemented via (2+1)D
viscous hydrodynamics with averaged initial conditions.
The parameter which controls the coupling between the jet and the medium, $\mu_\textrm{min}$,
is varied between  $1.3\pi T$ and $1.8 \pi T$, where $T$ is the temperature of the QGP
in the model, in the calculation shown here.
The choice of  $\mu_\textrm{min}$ values is motivated by comparisons with other
jet measurements.
These parameters have been shown~\cite{Ke:2020nsm} to provide
a reasonable description of measurements of the \RAA\
values of $B$- and $D$-mesons~\cite{CMS:2017uoy,CMS:2017qjw}.
The calculation by Dai et al.\
in Ref.~\cite{Dai:2018mhw} is based on a Langevin transport model
describing the evolution of \bquarks\ and their collisional energy loss and a
higher-twist description of radiative energy loss for both heavy and light partons.
This model also includes the dead-cone effect for
\bjets\ and also uses a (2+1)D viscous hydrodynamic medium with averaged initial conditions.
The parameter controlling the coupling of the jet to the medium, $q_0$, is set to be 1.2~$\GeV^2/$fm.
The LIDO model shows good agreement with the data for both inclusive jet and \bjet\ \RAA,
although with the inclusive jets on the low side and the \bjet\ case on the high side,
for all three centralities considering model and data uncertainties.
The calculation in Ref.~\cite{Dai:2018mhw} is below the measured \RAA\
in central and semi-central collisions for both \bjets\ and inclusive jets.
 
\begin{figure}[htbp]
\centering
\includegraphics[width=0.65\textwidth]{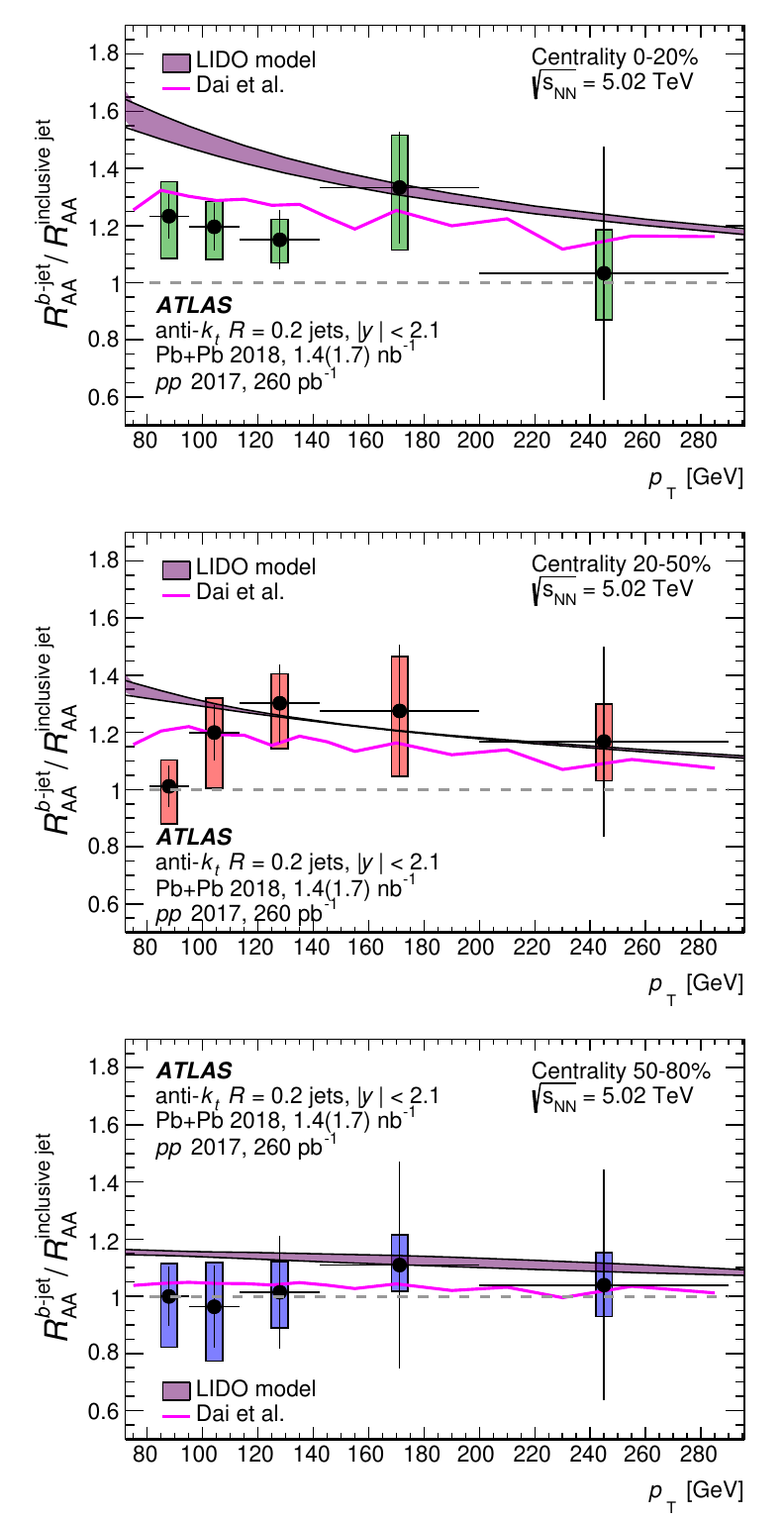}
\caption{Ratio of \bjet\ \RAA\ to the inclusive jet \RAA\ for each centrality class.
Ratios are compared with theory calculations~\cite{Ke:2018jem,Ke:2018tsh,Dai:2018mhw}.
The boxes represent the systematic uncertainties and the error bars
represent the statistical uncertainties.
For the LIDO calculation, the width of the band shows the variation of the $\mu_\textrm{min}$
parameter from $1.3\pi T$ (upper edge) to $1.8\pi T$ (lower edge).}
\label{fig:RAA_light_b_comp_ratio}
\end{figure}

In order to assess any difference between the \bjet\ and inclusive jet
\RAA, the ratio $\RAA^{\bjet} / \RAA^\textrm{inclusive jet}$ is presented in
Figure~\ref{fig:RAA_light_b_comp_ratio}
for each centrality class.
This provides a
more precise comparison of the \RAA\ values than in Figure~\ref{fig:RAA_light_b_comp}.
The results suggest that in central collisions the suppression of \bjets\ is less than
that of inclusive jets, with an overall significance of approximately $1.7\sigma$.
The calculation by Dai et al.~\cite{Dai:2018mhw} agrees well with the ratio in all centrality classes,
while LIDO model calculations tend to overestimate the double ratio at low \pt,
especially in central collisions, but agree well with the data at higher \pt.
In the LIDO model, the difference between the inclusive jet and \bjet\ \RAA\ values
is not expected to be entirely a result of the large mass of the \bquark,
and qualitatively similar differences are seen between the two \RAA\ values in that model
with and without the inclusion of the dead-cone effect~\cite{Ke:2020nsm}.
 
Differences in the internal structure of the \bjets, such as those that come from gluon-splitting processes,
are also expected to be important in determining the \bjet\ \RAA\ value.
Based on the \pythiaeight\ MC samples,
the fractions of jets which are initiated by gluons in the inclusive jet and \bjet\ samples have
opposite trends with \pt.  The fraction increases from 23\% to 45\% over the range of 80 to 250~\GeV\
for the \bjets\  and decreases from 61\% to 47\% over the same range for the inclusive jets.
 
These results are compatible with previous results from CMS~\cite{Chatrchyan:2013exa}  (although both the collision energy and jet radius are different) but have better precision.
Additional measurements of the fragmentation functions of \bjets\ in heavy-ion collisions would improve the uncertainties in future measurements using the method used in this article.
 
% End of text imported from the .//chapters/results.tex input file

\FloatBarrier

\section{Summary}
\label{sec:summary}

% The next lines are included from the .//chapters/conclusion.tex input file
This paper reports cross-sections for \bjets\ and inclusive jets in \pbpb\ and \pp\ collisions,
both at $\sqrt{s_{_\textrm{NN}}}=5.02$~\TeV, and recorded by the ATLAS detector at the LHC.
The measurement uses three datasets:
1.4~\invnb\ and 1.7~\invnb\ of \pbpb\ collisions collected in 2018 for \RTwo\ \bjets and inclusive jets respectively,
and 260~\invpb\ of \pp\ collisions collected in 2017 for \RTwo\ and \RFour\ \bjets\ and \RTwo\ inclusive jets.
The \bjet\ cross-section in \pp\ collisions is compared with a theoretical calculation
and Monte Carlo generator predictions.
The \bjet\ and inclusive jet per-event yields for \RTwo\ jets
and the corresponding nuclear modification factor, \RAA, are also reported for
\pbpb\ collisions at the same per-nucleon collision energy.
The \RAA\ values are found to decrease with increasing collision centrality
for both \bjets\ and inclusive jets.
In order to more directly compare the suppression of \bjets\ and inclusive jets,
the ratio of the \RAA\ values is presented. The central values of this ratio
suggest that the \RAA\ for \bjets\ is larger than that for inclusive jets
in central \pbpb\ collisions.
The observed differences may arise primarily from the different mixture of quark and gluon jets in the inclusive jets and $b$-jets,
and the $b$-quark mass effect may be subdominant in the kinematic range measured here.
However, the current systematic uncertainties do not permit a more definitive statement.
This highlights the need for more precise measurements of this quantity, mainly the necessity to measure the fragmentation functions of \bjets\ in heavy-ion collisions.
The measurements are compared with theoretical calculations and suggest a role for mass and colour-charge effects in partonic energy loss in heavy-ion collisions.

% End of text imported from the .//chapters/conclusion.tex input file

\section*{Acknowledgements}

% The next lines are included from the .//acknowledgements/Acknowledgements.tex input file

We thank CERN for the very successful operation of the LHC, as well as the
support staff from our institutions without whom ATLAS could not be
operated efficiently.
 
We acknowledge the support of
ANPCyT, Argentina;
YerPhI, Armenia;
ARC, Australia;
BMWFW and FWF, Austria;
ANAS, Azerbaijan;
CNPq and FAPESP, Brazil;
NSERC, NRC and CFI, Canada;
CERN;
ANID, Chile;
CAS, MOST and NSFC, China;
Minciencias, Colombia;
MEYS CR, Czech Republic;
DNRF and DNSRC, Denmark;
IN2P3-CNRS and CEA-DRF/IRFU, France;
SRNSFG, Georgia;
BMBF, HGF and MPG, Germany;
GSRI, Greece;
RGC and Hong Kong SAR, China;
ISF and Benoziyo Center, Israel;
INFN, Italy;
MEXT and JSPS, Japan;
CNRST, Morocco;
NWO, Netherlands;
RCN, Norway;
MEiN, Poland;
FCT, Portugal;
MNE/IFA, Romania;
MESTD, Serbia;
MSSR, Slovakia;
ARRS and MIZ\v{S}, Slovenia;
DSI/NRF, South Africa;
MICINN, Spain;
SRC and Wallenberg Foundation, Sweden;
SERI, SNSF and Cantons of Bern and Geneva, Switzerland;
MOST, Taiwan;
TENMAK, T\"urkiye;
STFC, United Kingdom;
DOE and NSF, United States of America.
In addition, individual groups and members have received support from
BCKDF, CANARIE, Compute Canada and CRC, Canada;
PRIMUS 21/SCI/017 and UNCE SCI/013, Czech Republic;
COST, ERC, ERDF, Horizon 2020 and Marie Sk{\l}odowska-Curie Actions, European Union;
Investissements d'Avenir Labex, Investissements d'Avenir Idex and ANR, France;
DFG and AvH Foundation, Germany;
Herakleitos, Thales and Aristeia programmes co-financed by EU-ESF and the Greek NSRF, Greece;
BSF-NSF and MINERVA, Israel;
Norwegian Financial Mechanism 2014-2021, Norway;
NCN and NAWA, Poland;
La Caixa Banking Foundation, CERCA Programme Generalitat de Catalunya and PROMETEO and GenT Programmes Generalitat Valenciana, Spain;
G\"{o}ran Gustafssons Stiftelse, Sweden;
The Royal Society and Leverhulme Trust, United Kingdom.
 
The crucial computing support from all WLCG partners is acknowledged gratefully, in particular from CERN, the ATLAS Tier-1 facilities at TRIUMF (Canada), NDGF (Denmark, Norway, Sweden), CC-IN2P3 (France), KIT/GridKA (Germany), INFN-CNAF (Italy), NL-T1 (Netherlands), PIC (Spain), ASGC (Taiwan), RAL (UK) and BNL (USA), the Tier-2 facilities worldwide and large non-WLCG resource providers. Major contributors of computing resources are listed in Ref.~\cite{ATL-SOFT-PUB-2021-003}.

% End of text imported from the .//acknowledgements/Acknowledgements.tex input file

\clearpage

\printbibliography

\clearpage
% ATLAS Collaboration author list
% Reference date of HION-2018-24 is 2022-02-04
% Author list last updated on date 13-JUN-23
% Data extracted on 13-Jun-2023 for paper reference HION-2018-24
% at 6:20pm
 
\begin{flushleft}
\hypersetup{urlcolor=black}
{\Large The ATLAS Collaboration}

\bigskip

\AtlasOrcid[0000-0002-6665-4934]{G.~Aad}$^\textrm{\scriptsize 101}$,
\AtlasOrcid[0000-0002-5888-2734]{B.~Abbott}$^\textrm{\scriptsize 119}$,
\AtlasOrcid[0000-0002-7248-3203]{D.C.~Abbott}$^\textrm{\scriptsize 102}$,
\AtlasOrcid[0000-0002-1002-1652]{K.~Abeling}$^\textrm{\scriptsize 55}$,
\AtlasOrcid[0000-0002-8496-9294]{S.H.~Abidi}$^\textrm{\scriptsize 29}$,
\AtlasOrcid[0000-0002-9987-2292]{A.~Aboulhorma}$^\textrm{\scriptsize 35e}$,
\AtlasOrcid[0000-0001-5329-6640]{H.~Abramowicz}$^\textrm{\scriptsize 150}$,
\AtlasOrcid[0000-0002-1599-2896]{H.~Abreu}$^\textrm{\scriptsize 149}$,
\AtlasOrcid[0000-0003-0403-3697]{Y.~Abulaiti}$^\textrm{\scriptsize 116}$,
\AtlasOrcid[0000-0003-0762-7204]{A.C.~Abusleme~Hoffman}$^\textrm{\scriptsize 136a}$
% ...
\bigskip
\\

% ...
$^{29}$Physics Department, Brookhaven National Laboratory, Upton NY; United States of America.\\
% ...
$^{35}$$^{(a)}$Facult\'e des Sciences Ain Chock, R\'eseau Universitaire de Physique des Hautes Energies - Universit\'e Hassan II, Casablanca;$^{(b)}$Facult\'{e} des Sciences, Universit\'{e} Ibn-Tofail, K\'{e}nitra;$^{(c)}$Facult\'e des Sciences Semlalia, Universit\'e Cadi Ayyad, LPHEA-Marrakech;$^{(d)}$LPMR, Facult\'e des Sciences, Universit\'e Mohamed Premier, Oujda;$^{(e)}$Facult\'e des sciences, Universit\'e Mohammed V, Rabat;$^{(f)}$Institute of Applied Physics, Mohammed VI Polytechnic University, Ben Guerir; Morocco.\\
% ...
$^{55}$II. Physikalisches Institut, Georg-August-Universit\"{a}t G\"ottingen, G\"ottingen; Germany.\\
% ...
$^{101}$CPPM, Aix-Marseille Universit\'e, CNRS/IN2P3, Marseille; France.\\
$^{102}$Department of Physics, University of Massachusetts, Amherst MA; United States of America.\\
% ...
$^{116}$Department of Physics, New York University, New York NY; United States of America.\\
% ...
$^{119}$Homer L. Dodge Department of Physics and Astronomy, University of Oklahoma, Norman OK; United States of America.\\
% ...
$^{136}$$^{(a)}$Departamento de F\'isica, Pontificia Universidad Cat\'olica de Chile, Santiago;$^{(b)}$Millennium Institute for Subatomic physics at high energy frontier (SAPHIR), Santiago;$^{(c)}$Instituto de Investigaci\'on Multidisciplinario en Ciencia y Tecnolog\'ia, y Departamento de F\'isica, Universidad de La Serena;$^{(d)}$Universidad Andres Bello, Department of Physics, Santiago;$^{(e)}$Instituto de Alta Investigaci\'on, Universidad de Tarapac\'a, Arica;$^{(f)}$Departamento de F\'isica, Universidad T\'ecnica Federico Santa Mar\'ia, Valpara\'iso; Chile.\\
% ...
$^{149}$Department of Physics, Technion, Israel Institute of Technology, Haifa; Israel.\\
$^{150}$Raymond and Beverly Sackler School of Physics and Astronomy, Tel Aviv University, Tel Aviv; Israel.\\
% ...
$^{a}$ Also Affiliated with an institute covered by a cooperation agreement with CERN.\\
$^{b}$ Also at Borough of Manhattan Community College, City University of New York, New York NY; United States of America.\\
$^{c}$ Also at Bruno Kessler Foundation, Trento; Italy.\\
$^{d}$ Also at Center for High Energy Physics, Peking University; China.\\
$^{e}$ Also at Centro Studi e Ricerche Enrico Fermi; Italy.\\
$^{f}$ Also at CERN, Geneva; Switzerland.\\
% ...

\end{flushleft}

% Created with Glance <Atlas.Glance@cern.ch>

\end{document}